\documentclass[12pt,preprint,a4paper]{aastex}
\usepackage{float}
\usepackage{amsmath}
\usepackage{natbib}
\usepackage[english]{babel}
\usepackage{footnote}
\usepackage{pdflscape}

\defcitealias{2012AA...539A.143N}{NP12}
\newcommand{\rion}[2]{{\ensuremath{\mbox{\rm #1$\,${\sc\expandafter{\romannumeral#2\relax}}}}}}

\begin{document}

\title{Abundance scaling in stars, nebulae and galaxies}
\shorttitle{Abundance scaling}
\shortauthors{Nicholls et~al.}

\author{ David C. Nicholls\altaffilmark{1}, Ralph S. Sutherland\altaffilmark{1}, Michael A. Dopita\altaffilmark{1}, Lisa J. Kewley\altaffilmark{1}, \& Brent A. Groves\altaffilmark{1},}
\email{david.nicholls@anu.edu.au}
\altaffiltext{1}{Research School of Astronomy and Astrophysics, Australian National University, Canberra, ACT, Australia }

\begin{abstract}
We present a new basis for scaling abundances with total metallicity in nebular photoionisation models, based on extensive Milky Way stellar abundance data, to replace the uniform scaling normally used in the analysis of \rion{H}{2} regions. Our goal is to provide a single scaling method and local abundance reference standard for use in nebular modelling and its key inputs, the stellar atmosphere and evolutionary track models. We introduce a parametric enrichment factor, $\zeta$, to describe how atomic abundances scale with total abundance, and which allows for a simple conversion between scales based on different reference elements (usually oxygen or iron) . The models and parametric description provide a more physically realistic approach than simple uniform abundance scaling. With appropriate parameters, the methods described here may be applied to \rion{H}{2} regions in the Milky Way, large and dwarf galaxies in the local universe, Active Galactic Nuclei (AGNs), and to star forming regions at high redshift.
\end{abstract}


\footnotesize{Accepted by MNRAS 11 Dec 2016}

\section{Introduction}
Photoionisation modelling provides a powerful tool for understanding the physics of \rion{H}{2} regions and other ionised nebulae. Its aim is to solve the radiative transfer equations and associated physics as the ionising radiation from the central star cluster traverses the gas. The goal is to predict the emission spectra arising from the ionised nebula, as a basis for interpreting observations, allowing us to determine the physical conditions and chemical abundances in the nebula.

There are a number of problems in developing reliable photoionisation models. First, a full knowledge of the atomic data for elements involved in the processes is critical. This has improved a great deal in recent years through resources such as the CHIANTI Database \citep{1997AAS..125..149D, 2013ApJ...763...86L}, but there are still gaps in our knowledge. However, the situation is likely to improve with time, and our current knowledge is sufficient to construct physically realistic photoionisation models of nebulae.

Second, our ability to synthesise accurately the spectrum of the source of the ionising radiation is far from complete. In \rion{H}{2} regions, this involves the estimation of the emission spectrum of the stars at the heart of the region. This depends on three factors: reliable models of stellar atmospheres ranging from hot O-stars to cool dwarfs; the evolutionary tracks of the stars as they age, for the lifetime of the ionised region, typically 10 to 20 Myr; and effective synthesis of the total spectrum from ensembles of stellar populations.

Worse, some of the commonly used stellar atmosphere model sets cover only sparsely (if at all) the hotter stars that dominate the excitation of \rion{H}{2} regions. The stellar evolutionary tracks generally are not well matched to the modelled main sequence stars, and use different metallicity standards, in part because these standards have been based on solar abundance references that have changed with time.

Population synthesis models combine the stellar models and evolutionary tracks, but cannot generate physically realistic results without consistent input data.  If we hope to build physically realistic photoionisation models, we need a single metallicity standard for future work on stellar atmosphere models, evolutionary track models and nebular models, to put them on the same footing. The lack of a consistent abundance scale introduces errors into nebular photoionisation models and makes the works of different authors difficult to compare.

Third, abundance scaling of elements at metallicities lower than the reference standard has been well explored in stars, but in the nebular modelling community only the simplest uniform scaling assumptions, or arbitrary adjustments to these, appear to have been used. Stellar astronomers have known for a long time, for example, that iron abundances relative to $\alpha$-element abundances have changed both over time and with the galactic  environment since the formation of the earliest stars \citep[e.g.,][]{1993ASPC...48..727W}. Consequently, photoionisation models for nebulae with different metallicities need to take this variation into account, and in general this has not been done.

Fourth, the extent of element depletion into dust in \rion{H}{2} regions and in different galactic environments is poorly known. It is possible to measure dust composition in the interstellar medium through absorption of starlight, but this does not tell us much about the nature of the dust in the giant molecular clouds from which \rion{H}{2} regions form, nor how the dust is destroyed by the ionising radiation in nebulae. Dust depletion can be measured, for example for refractory elements, by comparing nebular abundances with the photosphere abundances of the central stars (where available). A third approach, using photoionisation models, allows us to estimate dust depletion through comparison of models and observations. This paper addresses the second and third of these problems, and outlines how we propose to tackle the fourth.

We adopt a standard, present-day scale, extended from the {\em cosmic abundance standard} developed by \cite{2012AA...539A.143N}, based on the observed metallicities of 29 main-sequence B-stars in local galactic region, augmented with data from other recent sources for elements which are of minor importance in nebular and stellar modelling.  This is a local, present-day scale, rather than the conventional solar scale(s), where the abundance values include minor evolutionary effects overlaid on a scale deriving from the proto-solar nebula from ~5 Gyr ago, and uncertainties with the origin of the proto-solar nebula.  To avoid confusion with the Universe at large, we refer to the extended scale, {\em together with the associated scaling behaviour}, as the ``Galactic Concordance''. We suggest that this reference standard and scaling system be used for consistency in stellar atmosphere modelling, stellar evolutionary tracks and nebular models.

We examine stellar metallicity data assembled over the past two decades, from Milky Way observations and nearby dwarf galaxies, to derive a model for the way individual element abundances scale with total nebular metallicity. We suggest this should replace the simple linear scaling used in models to date.  Using piece-wise linear fits to the stellar abundance data, we show that abundances expressed in the iron-based stellar metallicity scale can be readily converted to the oxygen-based nebular metallicity scale, and vice-versa. We use the stellar data available to derive general rules for the nebular scaling of the important nebular elements as a function of total nebular oxygen metallicity.

Finally we suggest approaches to estimate nebular dust depletion, including comparison of nebular photoionisation models with the observed emission line data from simply-structured (and thus reliably modelled) \rion{H}{2} regions.

\section{The need for a consistent standard abundance set and a realistic scaling model}

\subsection{The problem of inconsistency}

Full radiative transfer nebular modelling requires inputs from a number of sources.  Among these, an accurate estimate of the central star cluster excitation spectrum is critical to to production of realistic nebular models. This requires detailed stellar atmosphere and stellar evolutionary track models.  These are convolved through a spectrum synthesis application which takes the raw stellar spectra and tracks, and builds a composite spectrum based on estimates of the ionising cluster initial mass function and the cluster evolution with time.

To generate a realistic  cluster spectrum, grids of stellar models are needed, and the spectral paths taken by the stars as they evolve.  There is a dearth of such models.  A major concern is that what models there are are based on abundance standards from different eras. For example, the WMBasic atmosphere models \citep[][and earlier papers]{2012AA...538A..75P} are based on the \citet{1989GeCoA..53..197A} solar photospheric standard abundances whereas the Geneva evolutionary track models \citep[][and other papers in this sequence]{2012AA...537A.146E} are based on solar photospheric abundances from \citet{2005ASPC..336...25A}. The oxygen abundance differs between these sources by 0.27 dex and the iron abundance by 0.22 dex.   Such differences cast doubt on the reliability of combining the track and atmosphere data as inputs to population synthesis applications.

In order to generate plausible nebular models, it is important that the stellar atmospheres and evolutionary tracks be computed using a common abundance standard.  In addition, the manner in which the abundances scale relative to each other is critical for modelling atmospheres, tracks and nebulae at metallicities less than the reference standard (e.g., ``solar'').

\subsection{The effects of abundance scaling on line diagnostics}
\begin{figure}
\includegraphics[width=\columnwidth]{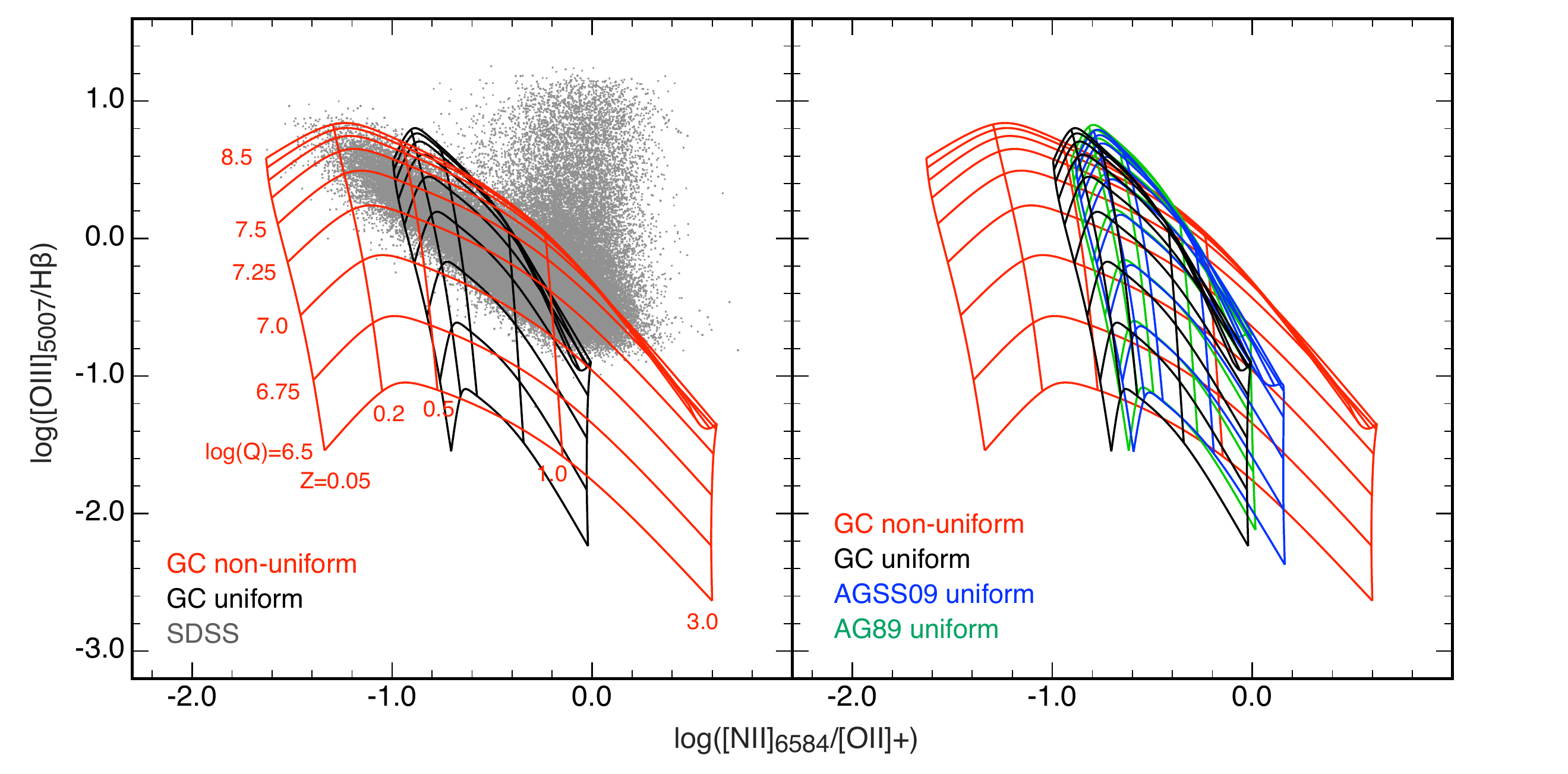}
     \caption{An illustration of the importance of non-uniform scaling in nebular modelling. The left panel shows grid plots for the strong line ratios log[\rion{O}{3}]$_{5007}$/H$\alpha$ vs. log([\rion{N}{2}]$_{6584}$/[\rion{O}{2}]$_{3726+3729}$), showing the ionisation parameter log(Q) (left to right) vs. metallicity Z relative the the standard value (near vertical lines), using the non-uniform scaling  and standard abundances proposed in this paper (red), and the same metallicity  standard and scaled values but with uniform scaling (black). The grey points are data for objects where \rion{O}{3} 4363\AA\ is detected, from the Sloan Survey Data Release 7 \citep{2009ApJS..182..543A}. They show clearly that uniform scaling is not capable of reproducing the observed data, while non-uniform scaling fits the observations well. (The vertical spray of SDSS points arises from active galactic nuclei, not modelled here). The right panel shows the same grids, but with uniform scaling using three different metallicity standards from \citet{1989GeCoA..53..197A, 2009ARAA..47..481A} and the GC scale from this work, green, blue and black, respectively. The different standard scales cause other problems of compatibility with stellar atmosphere and evolutionary tracks, but do not greatly affect the photoionisation model grids, and none can fit the observations using uniform scaling.
     }
\label{fig1}
\end{figure}

We use the {\sc Mappings} photoionisation modelling code \citep{2013ApJS..208...10D} to generate nebular strong line ratio grids for the ionisation parameter log(Q)\footnote{The ionisation parameter, Q, is the velocity of the ionisation front that the radiation field can drive through through medium. It is the ionising photon flux in photons cm$^{-2}$ s$^{-1}$ divided by the neutral hydrogen number density in cm$^{-3}$, and an indicator of the physical conditions at the inner edge of the photoionised zone in an \rion{H}{2} region.} vs. oxygen metallicity, the inadequacy of simple uniform abundance scaling (i.e., the same ratio for all elements) is apparent.  Uniform scaling is incapable of explaining the observations. Conversely the simple non-uniform scaling model we describe below does a very good job of matching observational data. Figure \ref{fig1} demonstrates this.

 While the geometry adopted for an \rion{H}{2} region model plays a key role in the predicted emission line outputs, we find that a plane parallel geometry provides a computationally tractable result that matches observations well. In this example, we assume constant pressure conditions with log(P/k) = 6.0, where P is the pressure and k is the Boltzmann constant. We adopt the WMBasic stellar atmosphere models \citep{2003ApJ...599.1333S}, the Geneva evolutionary tracks for rotating stars \citep{2012AA...537A.146E}, and use Starburst99 with a Salpeter IMF \citep{2014ApJS..212...14L} and continuous stellar evolution sampled at 5 Myr. For atomic abundances we use the Galactic Concordance scale, described in detail below.

Figure \ref{fig1} (left panel) shows the strong line ratios log[\rion{O}{3}]$_{5007}$/H$\alpha$ plotted vs. log([\rion{N}{2}]$_{6584}$/[\rion{O}{2}]$_{3726+3729}$), using the non-uniform scaling described below and standard GC abundances (red), and the same metallicity  standard but with uniform scaling (black). The grey points are data from the Sloan Survey Data Release 7 \citep{2009ApJS..182..543A}. They show that uniform scaling is not capable of reproducing the observed data, while non-uniform scaling fits the observations well. (The vertical spray of SDSS points arises from active galactic nuclei, not modelled here). Other strong line ratios also show discrepancies between models and observations when uniform scaling is used..

The choice of a ``standard'' abundance reference does not play a major role in explaining the observational data. Figure \ref{fig1} (right panel) shows the same grids using three ``standard'' abundance sets with uniform scaling.  None of them can explain the observational data.

\subsubsection{Differences arising from different abundance standards}

It is useful to explore what differences occur in metallicity and ionisation parameter, log(Q), using different abundance references. The grids in Figure \ref{fig1} (right panel) provide a guide to these differences.  For uniform abundance scaling, we can estimate the effects of using different abundance reference sets (black,  Galactic Concordance (this work); green \citet{1989GeCoA..53..197A} and blue \citet{2009ARAA..47..481A}.  The metallicity differences for this combination of line ratios vary between 0 and 0.15 at low metallicity and between 0 and 1.4 for metallicities $>$ 2.5.  Ionisation parameter differences are generally log(Q) $<$ 0.15.

Figure \ref{fig1} shows that the choice of abundance reference is less important to model outputs than the manner in which the scaling is calculated. However, whatever standard is used, it is vital that all inputs to nebular models (stellar atmospheres and evolutionary tracks, and nebular parameters) use the same abundance reference.

\section{Abundance scaling}

\subsection{The ``solar standard''}

The choice of a standard metallicity scale plays an important role in nebular stellar atmosphere and evolutionary track models.  In the past the only detailed reference values have been the solar  abundances. This has advantages because the abundances of many elements have been accurately measured in the solar photosphere. But it has significant shortcomings: some elements (for example, F, Cl, Ne and Ar) have not been directly detected, or only marginally detected, in the solar atmosphere; some elements (for example, He and Li) have been processed in the sun during its evolution and photospheric abundances do not necessarily reflect bulk or proto-solar values; and some elements that are crucial for nebular physics (i.e., oxygen) are difficult to measure in the solar spectrum \citep[see discussion in][]{2009ARAA..47..481A}. As a result, measured solar abundance values for some elements have differed considerably over the past 40 years. They also generate three ``standard'' abundance sets: photospheric, bulk and proto-solar nebular.  The published solar abundance estimates are listed in Appendix Table \ref{tablea1}. Of major concern is that the measured solar oxygen abundance value has varied by a factor of 1.9, iron by 1.7, nitrogen by 1.9, and carbon by 2.1. Further, solar photospheric values differ from the calculated bulk and proto-solar values.  Different workers have used different standards for comparison. In general, the latest solar values have been used, but this can lead to confusion when comparing data from earlier works based on previous solar standards.

\subsection{Relative scaling of elements}

The solar standard reference (or any alternative standard) should, ideally, specify the relative abundances for local, present day values. This presents a further problem for the solar values, as they derive from proto-solar nebula values from $\sim$5 Gyr ago, where the sun formed. What is not specified in the solar values is how element abundances scale relative to one another over time, from the early Universe to the present day. If we are to undertake photoionisation modelling at early epochs, or in systems with different star formation histories than the Milky Way, we need to understand how the relative abundances of each element vary with total metallicity through time.

In nebular modelling, a simple scaling of total metallicity by the same multiplicative  factor has been the norm for all but a few elements (such as He, C and N).   This was an acceptable approximation in the absence of any better information, but the assumption of uniform scaling has a significant impact on models of the physics of \rion{H}{2} regions, justifying a careful examination of what we may be able to determine about the actual scaling behaviour. See section 2 above.

The problems arise due to the rate at which the heavier elements have been formed in stars since early epochs. Different enrichment processes occur for the elements that are most important to the heating/cooling balance in \rion{H}{2} regions, so it is important to improve our estimates of actual scaling behaviour, insofar as this can be  determined.

 We also strike a definition problem: stellar abundances are measured relative to iron as the reference, whereas nebular abundances are measured relative to oxygen.  The ratios of these two reference elements have varied considerably with time: oxygen is principally produced in core-collapse supernovae (SN), whereas iron is principally produced in detonation SN. Core-collapse SN began enriching the primordial interstellar gas very early in the history of the Universe, but detonation SN have a delayed onset. So iron is relatively scarce early and increases rapidly after the delayed onset of detonation SN. Nitrogen is more complicated still: it is produced in both type of SN, but also in evolved stars, for example on the AGB branch and in hot young WN stars---some of these processes are prompt and some delayed, and some dependent on total metallicity and stellar mass.

Helium presents further problems. It is the second most common element in the Universe and is important in the nebular heating and cooling balance, especially at low total metallicities. It was created in the Big Bang and is being created continuously in stars, therefore its abundance has increased in the interstellar medium (ISM) from the primordial value to the present day value.  We have no detailed data on the historical rate of production of helium, but it is no doubt determined by the overall history of star formation, so, in the absence of better data, we assume a linear behaviour with oxygen metallicity. The calculation of helium abundance at different total metallicities depends on our estimate of the primordial value, itself prone to revision.

\subsection{An alternative reference standard}

Before attempting to calibrate the way abundances scale with total metallicity, we need an abundance reference standard set. As noted above, inconsistencies in stellar atmosphere and evolutionary track models can arise whenever a new standard solar abundance set is published. For this reason we adopt the ``Cosmic Abundance Standard'' proposed by \cite[][hereafter, NP12]{2012AA...539A.143N}, based on local B stars as the standard for the present day and the local region of the Milky Way, extended for completeness to include elements not present in NP12.

The suitability of the local B star abundances as a standard reference scale has been discussed in detail by NP12. Although the values derived from local B stars have uncertainties similar to the solar values, and are also potentially subject to revision if a more extensive stellar population is used, our reasons for proposing this set as a reference standard are: B star photospheric abundances measure the bulk abundances of the nebulae in which they recently formed; they formed locally in the Milky Way; and they provide an ensemble average over 29 stars, rather than depending on a single star (the sun) which may or may not be typical of current local abundances.

The NP12 list includes the eight most important nebular elements, He, C, N, O, Ne, Mg, Si, and Fe. We augment this set for completeness with the best estimates of abundances for elements not included in the NP12 list, for example from the most recent solar \citep{2015AA...573A..27G, 2015AA...573A..26S} and meteoritic abundances \citep{2015AA...573A..25S, 2015AA...573A..26S, 2015AA...573A..27G, 2009LanB...4B...44L}. This approach has been used in all previous solar reference lists. However, these other elements are less important in the physics of nebular processes. Although the values we suggest are not arbitrary, other sources could be chosen. We discuss the reasons for adopting the particular values in detail below.

The ``Galactic Concordance'' scale is given in Table \ref{table1}. It is encouraging to note that the local B star abundance for oxygen (12+log(O/H)=8.76) is closer to the estimated primordial solar abundance (8.73) from \citet{2009ARAA..47..481A} than to the often used solar photospheric abundance (8.69). The Galactic Concordance scale also includes scaling behaviour, as we now discuss.

\begin{table}
\small
\centering
\caption{Galactic Concordance abundances}
\label{table1}
\begin{tabular}{llrr} 
\hline \\
Z & X & 12+log(X/H) & Source\\
\hline
1 & H & 12.000 & -\\
2 & He & 10.990 & 1\\
3 & Li & 3.278 & 2\\ 
4 & Be & 1.320 & 2\\
5 & B & 2.807 & 2\\
6 & C & 8.423 & 3\\
7 & N & 7.790 & 4\\
8 & O & 8.760 & 1\\
9 & F & 4.440 & 2\\
10 & Ne & 8.090 & 1\\
11 & Na & 6.210 & 5\\
12 & Mg & 7.560 & 1\\
13 & Al & 6.430 & 5\\
14 & Si & 7.500 & 1\\
15 & P & 5.410 & 5\\
16 & S & 7.120 & 5\\
17 & Cl & 5.250 & 2\\
18 & Ar & 6.400 & 5\\
19 & K & 5.040 & 5\\
20 & Ca & 6.320 & 5\\
21 & Sc & 3.160 & 5\\
22 & Ti & 4.930 & 5\\
23 & V & 3.890 & 5\\
24 & Cr & 5.620 & 5\\
25 & Mn & 5.420 & 5\\
26 & Fe & 7.520 & 1\\
27 & Co & 4.930 & 5\\
28 & Ni & 6.200 & 5\\
29 & Cu & 4.180 & 5\\
30 & Zn & 4.560 & 5\\
\hline
Source &  \multicolumn{3}{l}{Reference} \\
\hline
1 & \multicolumn{3}{l}{\cite{2012AA...539A.143N}} \\
2 & \multicolumn{3}{l}{\cite{2009LanB...4B...44L}} \\
3 & \multicolumn{3}{l}{fit to O and Fe} \\
4 & \multicolumn{3}{l}{fit to N/O vs O curve} \\
5 & \multicolumn{3}{l}{\cite{2015AA...573A..25S, 2015AA...573A..26S},} \\
 & \multicolumn{3}{l}{\ \ \cite{2015AA...573A..27G}} \\
 \hline
\end{tabular}
\end{table}

\subsection{A new ``scaling parameter''}

Through long usage, the ``Z'' symbol specifies metallicity (that is, the combined mass fraction of all elements heavier than He) and Z$_0$ is the solar reference metallicity. Using the terminology ``Z/Z$_0$" as the measurement of metallicity relative to solar metallicity is ambiguous.  The ambiguity arises from the fact that whereas Z is strictly defined as the total abundance of elements heavier than helium (specified in the X(H)+Y(He)+Z = 1 mass fraction relationship), it has come to be loosely used in nebular analysis to mean the abundance of oxygen.  While oxygen makes the dominant contribution to total Z, the two values are not identical.

An alternative ``solar-independent scaling factor" is desirable, to avoid this ambiguity.  In order to put abundance scaling on a more systematic basis, we propose a scaling parameter, $\zeta$. The scaling origin standard fiducial point, $\zeta$ = 1 ($\log \zeta= 0$), is based on the mean values of local region B stars from \citet{2012AA...539A.143N}, and therefore refers to the present day chemical abundances in the local region of the Milky Way.

We also need to specify, with  $\zeta$, the element used as the scale reference. We could use any element or group of elements as the basis for scaling. The obvious choices are iron (used in the stellar abundance scale) or oxygen (employed in nebular physics), or the total metallicity, Z. We refer to these as $\zeta_{Fe}$, $\zeta_{O}$ and $\zeta_{Z}$, respectively. In practice, Fe and O are the most useful and the most widely used. While these scaling factors are different, we will show that they can be readily converted one to the other, using observed scaling behaviour derived from the stellar spectra (section 4).

\subsection{Stellar data as a guide to nebular scaling at low metallicities}

For \rion{H}{2} regions, the measured spread of metallicity ranges over $1/50 \lesssim Z/Z_0 \lesssim 2$, with most being found at $Z/Zo > 0.1$.  Stellar metallicities (measured on the [Fe/H] scale) span a wider range, between $\sim$10$^{-5}$ and $\sim$3 ``solar'' \citep[see][and subsequent papers in the series]{2013ApJ...762...28N}. It is not clear how relative nebular abundances of Fe and O scale in \rion{H}{2} regions at low abundances, since there is very little observational material on nebular Fe abundances, and we have little knowledge on how depletion onto dust affects the gas-phase abundance of Fe. This is especially true at low metallicities.

To deal with this problem, we can draw on the extensive information from stellar spectra as a guide to relative abundances of many elements at low metallicities, \citep[see, for example,][]{2013AA...552A...6G}. If we use data for main sequence stars, before their atmospheres have evolved due to local nucleosynthesis and dredge-up, we have useful information on the abundances in the \rion{H}{2} regions in which they formed, spanning a far greater range of total metallicity than possible from nebular data.

Major benefits of this approach are in the wide metallicity range available and the ability to avoid abundance uncertainties due to dust depletion. Minor problems include the different, Fe-based, scaling used in stellar measurements, and the difficulty of measuring the abundances of some elements, notably oxygen.  As we will show, the choice of scaling reference element is not a problem using the $\zeta$ parameter, and sufficient stellar spectra are available to provide an accurate idea of how the available stellar data for oxygen scales with iron.

In the following section, we explore the stellar data available for 6 elements that play a key role in controlling the nebular emission line spectrum (O, Ne, Mg,  Al, Si, S), and also Ca.  We present simple piece-wise linear fits to the abundances derived from the spectral data for these elements. We fit carbon using the iron scale so that it is consistent with the observed log(C/O) curves vs log(O/H) from \citet{2012AA...539A.143N}. For nitrogen a fit is obtained from stellar abundance log(N/O) values plotted vs stellar oxygen abundance, and similarly for Cl. From nebular measurements, we assume that the $\alpha$-process elements Ne and Ar scale with O directly. Stellar data for elements of minor nebular importance are treated similarly in Appendix A (Na, P, K, Sc, Ti, V, Cr, Mn, Co, Ni, Cu, Zn).

It is worth stressing that our purpose is not to model exactly how the abundances scale with iron, because of the intrinsic variation between individual stars and stellar populations even within the Milky Way, but to gain an overview of the trends where they are apparent, and to use these to build improved models for scaling nebular abundances.

\section{Abundance scaling in stellar data}

\subsection{Stellar data sources}

The study of the scaling of stellar abundance ratios has a long history. It was reviewed by \cite{1989ARAA..27..279W}, using the stellar data available at the time.  Since then, far more stellar data have become available, in terms of the stellar populations, the range of metallicities and the elements measured. For over two decades, much effort has been put into conducting surveys of stellar spectra \citep[see, for example][]{2003Msngr.114...20M, 2014AA...563A..44M}. In particular, work searching for evidence of extra-solar planets \citep[see, e.g.,][]{2013AA...552A...6G} has yielded an extensive collection of high quality stellar spectra. These provide direct measurements of how element abundances evolve with increasing total metallicity.

Some of these studies have been targeted at the derivation of abundances of specific elements, others have been directed at ranges of elements such as the iron-peak or the $\alpha$-elements.  Some investigate dwarf F, G and K stars in the solar neighbourhood, others concentrate on stars in the thin disk,  thick disk and halo. The stellar data references used here are listed in Table \ref{tablea2} and with each abundance plot (Figures \ref{fig2} - \ref{fig6} and \ref{figA1}- \ref{figA4}).  The stellar populations observed in these references are given in Table \ref{tablea3}. Taken as a whole, they provide an extensive database from which to explore abundance scaling over a wide range of metallicities.  We use data from these sources to develop models for the nebular scaling relations between each element.

In nebular modelling, the abundant elements which determine both the ionisation and thermal structures of nebulae, and hence the emission line spectra, are H, He, C, N, O, Ne, Mg, Si, Ar, and Fe. However, as the critical inputs to any nebular model include stellar atmospheres and evolutionary paths, we consider here all the elements up to Zn. Further, by including less abundant elements such as Ni and Cl, we make possible comparisons of observed line fluxes and model predictions, thereby allowing us to calibrate the model settings, and, ideally, estimate dust depletions.

The ease of measuring iron in stellar spectra makes it the natural choice for comparing with other elements. We therefore present the abundances of the elements considered here as a fraction of iron, vs iron, i.e., [X/Fe] vs. [Fe/H] (Equation \ref{eqn1}). Although it would be possible to plot [X/O] vs [O/H] from the stellar spectra, the number of stars with recorded oxygen is significantly less than those with iron measured in the spectrum, and the oxygen data is noisier than the iron data. Consequently we obtain fits to the abundance data for [Fe/H] scaling, and then invert them to oxygen as the scaling base, using the relation between [O/Fe] and [Fe/H].

\subsection{Scaling oxygen and the alpha-elements in stellar spectra}

Before looking at the stellar data element by element, a few general points are worth noting.

\begin{itemize}
\item A comparison of stellar abundance measurements using iron as the reference scale and the oxygen-scaled nebular measurements requires us to convert between the two scales. To do this we can observe how stellar oxygen abundance varies with stellar iron abundance, and use this to convert the observed stellar abundance scaling of the other elements between the iron and oxygen scales. The conversion is not a linear process, as different elements are synthesised at different rates during stellar evolution. This analysis is given in detail in the next section.

\item In general, the scaling of $\alpha$-elements with iron (log scale) in Milky Way stars is approximately constant for [Fe/H] $>$ -2.5 until the Type Ia supernovae begin to emerge, and then falls with increasing iron abundance starting at a well defined break point at [Fe/H] $\sim$ -1.0 \citep[see the discussion in][ especially their Figure 1]{1993ASPC...48..727W}.  The initial  abundance ratio of oxygen to iron is determined by the massive star IMF, and the break point is determined by the star formation rate. There is evidence that the massive star IMF is largely invariant \citep[see, for example,][]{1998ASPC..142...89W,2015AA...582A.122K}. Thus the stellar abundance scaling relations (Figures \ref{fig1} to \ref{figA4} below) remain largely constant for stars in the Milky Way. This may be a useful starting point for other large galaxies with similar evolutionary histories to the Milky Way, but may not be the case for smaller galaxies, or massive active starburst galaxies.

\item Stellar spectra analyses over the past decade \citep[e.g.,][]{2005AA...433..185B} provide evidence that different stellar populations in the thin disk, the thick disk, the bulge and the halo of the Milky Way have somewhat different star formation histories and can therefore be distinguished in abundance plots. As we wish to derive simple abundance scaling models as a guide to abundance scaling in nebulae, we base our models on ensemble average fits to the stellar scaling, rather than using single populations, although in most cases, the sampled populations are dominated by Milky Way thick disk stars. We provide a list of the sources used for each element, and the populations studied, in Table \ref{tablea2}. We list the populations studied in these sources in Table \ref{tablea3}.

\end{itemize}

\subsubsection{Oxygen abundances}
Oxygen is the most abundant element in \rion{H}{2} regions after H and He, and plays a dominant role in the physical processes. However, it is not an easy element to measure in stellar spectra, due to the weakness of the absorption lines, especially at low metallicity, and, in some cases, interference from adjacent nickel lines. As a consequence, there is significant  scatter in the computed abundance values.  The manner in which abundances are calculated is also critical: ignoring non-Local Thermodynamic Equilibrium (NLTE) effects can introduce errors of $\sim$0.1 dex in computed abundances even in the case of cool dwarf stars \citep{2013ApJ...764...78R}. 3D NLTE models give consistently better results than 1D or local thermodynamic equilibrium (LTE) methods.

Aggregate plots of oxygen abundance show considerable scatter, a result of different analysis methods and observational uncertainties. Figure \ref{fig2} (left panel) shows recent results from a 3D NLTE reanalysis of 648 stellar spectra from \citet{2015MNRAS.454L..11A} selected from high signal to noise ratio data sets.  We use the data from this work as they exhibit significantly less scatter than a simple aggregation of observations.

The data plotted are [O/Fe] vs [Fe/H] where these parameters are defined in the usual stellar formalism, in terms of the reference Fe and O values. The notation refers to a reference standard, usually a solar scale, but here we use it to refer to the Galactic Concordance scale , using number fractions rather than mass fractions:

\begin{equation}\label{eqn1}
[\rm{Fe/H}] = \log(Fe/H)_{star} - \log(Fe/H)_{reference}
\end{equation}
and
\begin{equation}\label{eqn2}
[\rm{O/Fe}] = \log(O/Fe)_{star} - \log(O/Fe)_{reference} \ .
\end{equation}

Each data set has been converted to the Galactic Concordance scale (see previous section).  The red lines are adopted fits to the observations, discussed below. They are close to the least-square fits to the data (see left panel, Figure \ref{fig1} ), but the scatter in the stellar data due to the different measured populations and intrinsic measurement and modelling uncertainties, make least-square fitting of little value.

In the reported stellar abundance data, there is increasing scatter below [Fe/H] $\sim$ -2.5. Some of this is due to measurement noise, and some is due to different methods used to derive the metallicity (LTE or NLTE, 1D or 3D). However, some scatter imay also be caused by intrinsic stochasticity in the abundance ratio, indicating stars that formed in regions where the elements in the ISM had not been uniformly enriched by sufficiently many core-collapse supernovae to generate a uniform abundance pattern \citep{1998ASPC..142...89W}. It is also possible that some scatter is due to the local influence of different types of core-collapse supernovae. More recent data for metal-poor stars from the thick disk from the RAVE survey \citep[Figure \ref{fig2},][dark green points, Mg and Si]{2011ApJ...737....9R} show a tighter spread than for the oxygen data between [Fe/H] = -2.8 and -1.0. To avoid uncertainty in deriving generic fits to the trends, we only attempt fits above [Fe/H] = -2.5.

For -2.5 $<$ [Fe/H] $<$ -1.0, there is a region where [O/Fe] appears approximately constant but with wide scatter. This is the zone where the enrichment by core-collapse supernovae has generated an ISM with approximately constant composition, and also where the element abundances are sufficient to reduce measurement uncertainty.

The next feature occurs at [Fe/H] = -1.0, where there is a breakpoint, followed by clear downward trend in [O/Fe] as Type 1a (detonation, or low-mass) supernovae commence enriching the ISM with large amounts of iron. The scatter in this region is likely due to the intrinsic diversity of stellar populations and individual evolutionary paths.

At some point, a further constant plateau at a lower value of [O/Fe] may be reached, when the enrichment balance of core-collapse and detonation supernovae again achieve a constant abundance ratio, assuming a continuing supply of interstellar gas. The stellar data do not show if, or exactly where this second breakpoint occurs, so, for the purposes of analysis we have set it at [Fe/H] = +0.5, corresponding to $\sim$3 times the current reference iron abundance. It does not seem likely that the enrichment of iron relative to other elements will continue indefinitely.   There is some evidence consistent with this assumption in the $\alpha$-element results presented by \citet[][in particular, their figure 19]{2011AA...530A.138C}.

A further point to note is that there appears to be a sharp upper limit, $\sim$+0.6, to the [Fe/H] values observed. The reason for this may be that no stars have yet formed above this iron abundance to enrich the ISM. It is the subject of a current research program (M. Asplund, 2016, pers. comm.). The equivalent widths of iron absorption lines are also difficult to measure at high metallicity against a stellar continuum eroded by numerous fine absorption lines.

\subsection{Stellar abundance scaling plots}

\subsubsection{Alpha-elements, oxygen, magnesium and silicon}
\begin{figure*}
	\includegraphics[width=\textwidth]{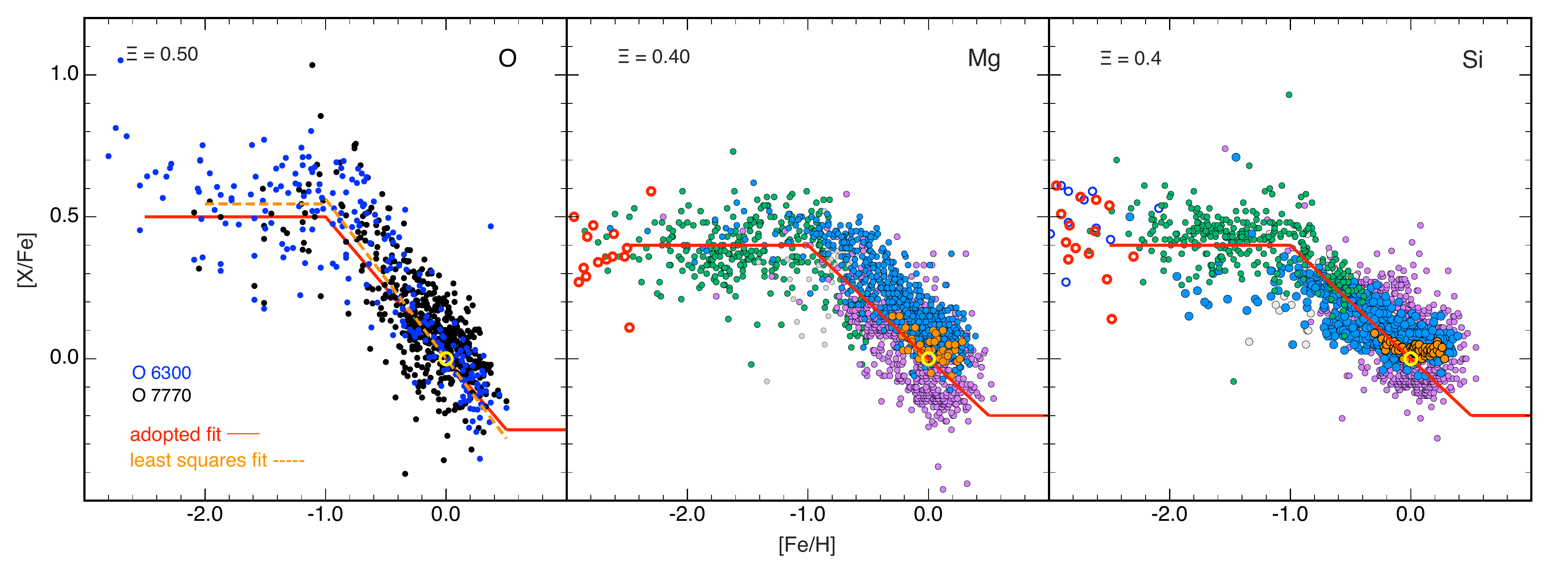}
    \caption{Scaling of O, Mg, Si vs. Fe from stellar spectra.
   Left panel:  Oxygen scaling as a function of [Fe/H] from \citet{ 2015MNRAS.454L..11A}, the most carefully and consistently reduced stellar oxygen data available. The adopted piece-wise linear fit is shown as a red line, and the standard (GC) metallicity (fiducial point) as a yellow circle. The dashed orange line is a piece-wise least-squares fit to the data, and differs from the adopted fit by far less than the intrinsic scatter of the stellar data.   The parameter $\Xi$ is defined in Equation \ref{eqn5} and specifies the low metallicity plateau value. Note that, in this and subsequent figures, while the trend lines drawn extend to [Fe/H] $<$ -2.5, we only use values $>$ -2.0 in our nebular fits.  Mid- and right panels show the stellar data for Mg and Si.
    Sources:
    \citep[][O, blue and black points]{ 2015MNRAS.454L..11A}
    \citep[][Mg, Si, dark green]{2011ApJ...737....9R},
    \citep[][Mg, Si, grey]{2012AA...545A..32A},
    \citep[][Mg, Si, orange]{2013AA...552A...6G},
    \citep[][Mg, Si, blue discs]{2014AA...562A..71B},
    \citep[][Mg, Si, purple]{2014AJ....148...54H},
    \citep[][Mg, Si, red circles]{2015Natur.527..484H}
     \citep[][Si, blue circles]{2004AA...416.1117C}}
    \label{fig2}
\end{figure*}

Figure \ref{fig2} (middle and right panels) shows the logarithmic abundance plots for the $\alpha$-elements magnesium and silicon, both of which exhibit the same behaviour as oxygen, repeated in this figure for comparison. It is clear that the upper breakpoint is very similar for each $\alpha$-element, at [Fe/H] $\sim$ -1.0 for the Milky Way stars analysed. This is consistent with the idea that for [Fe/H] $<$ -1.0, the abundance patterns are established by numerous core-collapse supernovae. Above that value, in each case, detonation supernovae commence the iron enrichment process.  Other similar published data, e.g., \citet[][their Figure 8]{2005AA...433..185B} show the same behaviour, but have not been included in the diagrams for clarity.

\subsubsection{Calcium, Sulphur and Aluminium}

\begin{figure*}
	\includegraphics[width=\textwidth]{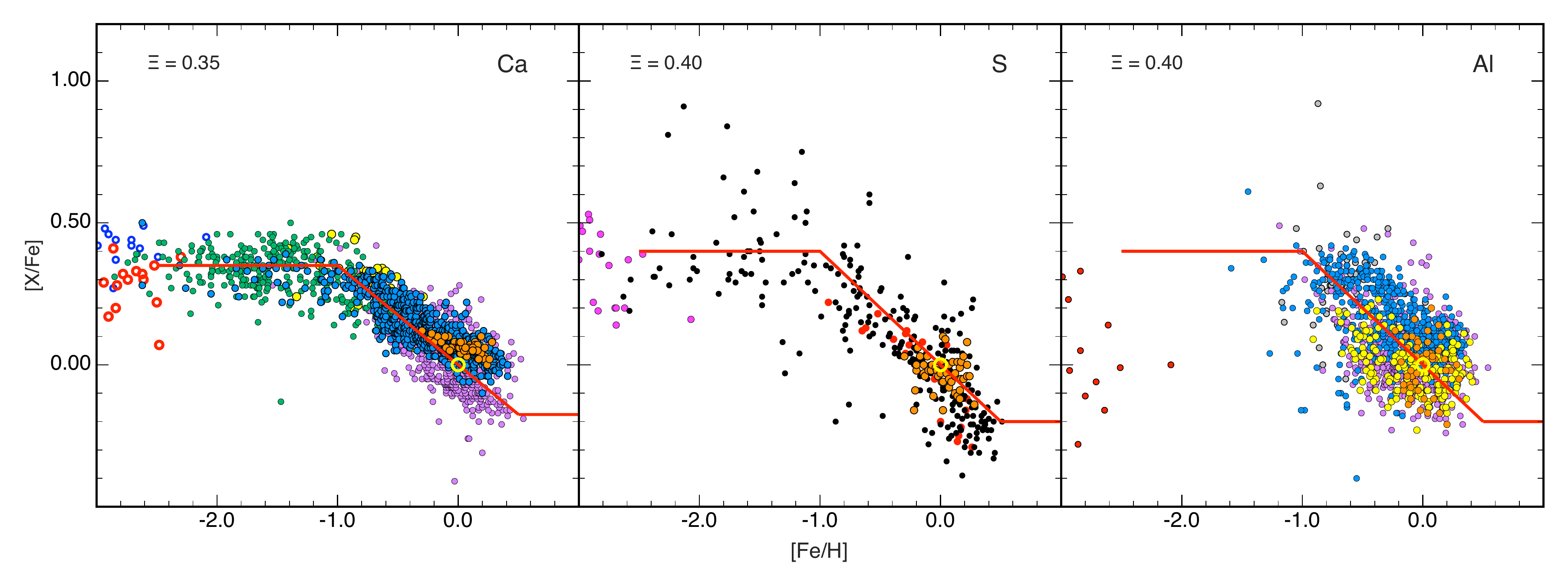}
    \caption{Scaling of Ca, S, Al vs. Fe from stellar spectra. Sources:
    \citep[][Ca, dark green]{2011ApJ...737....9R},
    \citep[][Ca, yellow]{2012AA...545A..32A},
    \citep[][Ca, red circles]{2015Natur.527..484H},
    \citep[][Ca, blue circles]{2004AA...416.1117C},
    \citep[][Ca, Al, purple]{2014AJ....148...54H},
    \citep[][Ca, Al, blue]{2014AA...562A..71B},
    \citep[][Ca, S, Al, orange]{2013AA...552A...6G},
    \citep[][S, black, red]{2005AA...441..533C, 2011AA...532A..98C},
    \citep[][S, magenta]{2011AA...528A...9S},
    \citep[][Al, yellow]{1993AA...275..101E},
    \citep[][Al, grey]{2010AA...509A..88A},
    \citep[][Al, red]{2004AA...416.1117C}}
    \label{fig3}
\end{figure*}

The data for Ca, S and Al in Figure \ref{fig3} follow oxygen, magnesium and silicon, as expected for elements generated by the alpha process. In the aluminium plot, the offset and scatter at low [Fe/H] may be the result of stellar model deficiencies in the older data, and so have not been included in the line fit.

\subsection{Scaling of other important nebular elements}

\subsubsection{Carbon and Nitrogen}

Carbon and nitrogen in nebulae present a problem when scaled to metallicities higher or lower than the standard baseline. \citet[][figure 4]{1993MNRAS.265..199V} showed that the nebular scaling of nitrogen with oxygen can be explained by a combination of primary nitrogen (a constant fraction of oxygen with increasing oxygen abundance) and secondary nitrogen (a linearly increasing fraction of oxygen with increasing oxygen in log space). The primary abundances originate from enrichment by core-collapse supernovae in the native gas cloud from which the \rion{H}{2} region formed, and the secondary abundances arise from delayed nucleosynthesis through hot-bottom burning and dredge-up in intermediate mass stars as they evolve.

The existence of primary nitrogen has been questioned for stellar spectra \citep{2005ARAA..43..481A}, but the nebular data consistently appear to follow the primary/secondary trend (see \citet{1993MNRAS.265..199V, 2000ApJ...542..224D, 2004ApJS..153....9G, 2013ApJS..208...10D}, and \citet{1999ApJ...511..639I}), so we accept this model as a useful description of the scaling behaviour of nitrogen. The observed dispersion in the values of log(N/O), especially at low metallicity, are discussed in \cite{2006AA...450..509G}. Clearly, this spread makes specifying a single description problematic, but we use the fit described here, noting that it should be treated as a starting point for modelling, rather than being prescriptive.

\begin{figure*}
	\includegraphics[width=\textwidth]{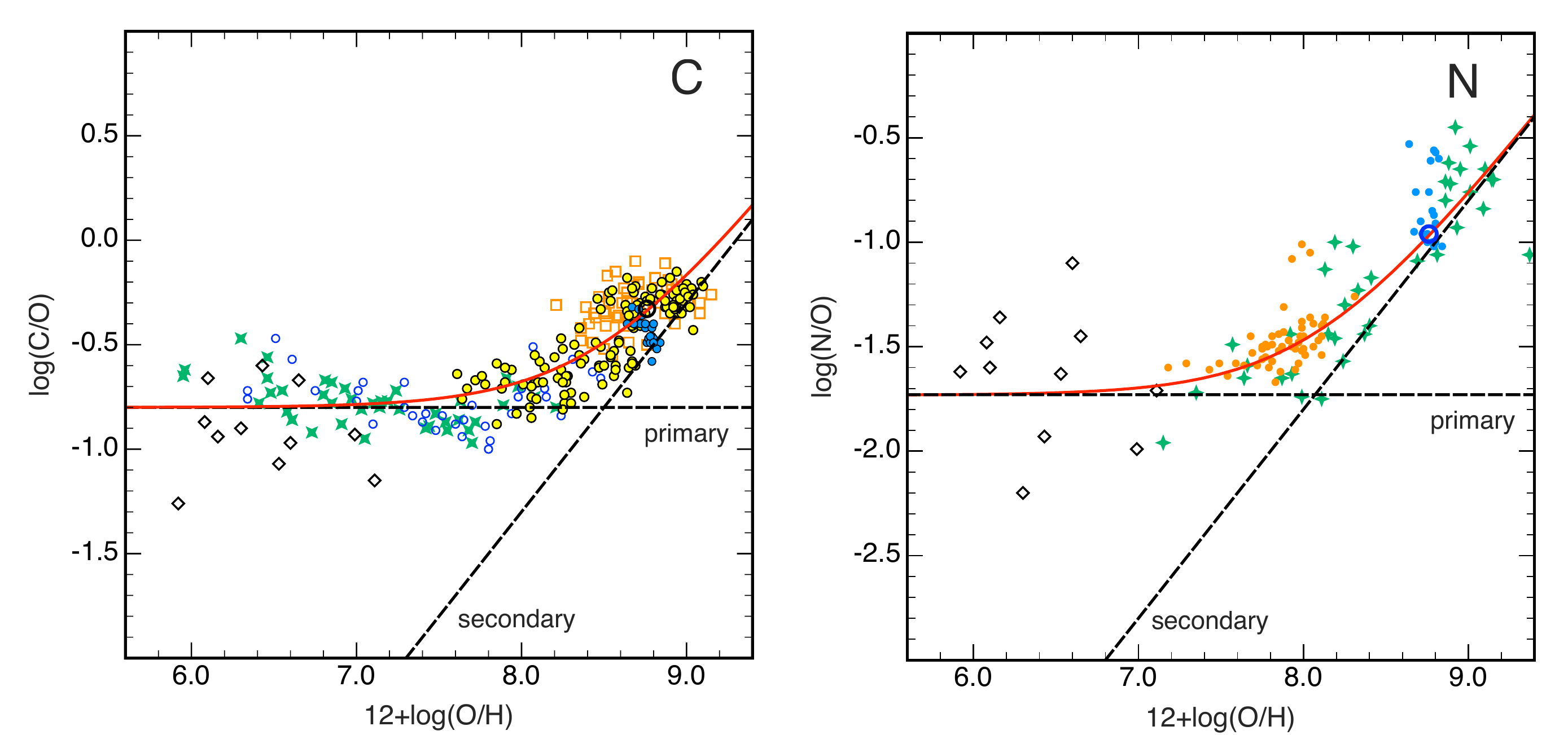}
    \caption{Scaling of C and N. C: sources:
    \citep[][orange squares]{1999AA...342..426G},
    \citep[][blue circles]{2004AA...414..931A},
    \citep[][green 'x']{2009AA...500.1143F},
    \citep[][black diamonds]{2005AA...430..655S},
    \citep[][blue discs]{2012AA...539A.143N},
    \citep[][yellow discs]{2014AA...568A..25N}
    N: sources:
    \citep[][orange discs]{1999ApJ...511..639I},
    \citep[][green '+']{2004AA...421..649I},
    \citep[][black diamonds]{2005AA...430..655S},
    \citep[][blue discs]{2012AA...539A.143N}}
    \label{fig4}
\end{figure*}

Figure \ref{fig4} presents plots of log(C/O) and log(N/O) vs. 12+log(O/H), showing the complex scaling behaviour due to primary and secondary sources. (We use the 12+log(O/H) scale, rather than [O/H], following the nebular physics convention). The left panel of that figure shows stellar carbon data from  \cite{1999AA...342..426G} (squares, galactic disk solar type dwarfs), \cite{2005AA...430..655S} (diamonds, halo metal-poor unmixed giants), \cite{2009AA...500.1143F} (crosses, halo solar type dwarfs and subgiants), and \citet{2012AA...539A.143N} (blue circles, B stars in the local region). The nitrogen data are also consistent with that from \citet{2006MNRAS.372.1069M, 2006AA...450..509G}. Likewise, the fits and data from \citet{2005AA...432..861G} are consistent with the carbon data presented here.

For carbon, the data from \cite{2004AA...414..931A} and  \cite{2014AA...568A..25N} were not used in the fit, as information on computation methods used (solar standard, N/LTE) was not clear, but the data from those sources suggests they are consistent with the fit. The scaling we assume for carbon depends on stellar observations with consequent caveats on their reliability. It is well established that some very early stars are rich in carbon while being very metal poor (carbon enhanced metal poor, or CEMP) \citep[e.g.][]{2013ApJ...762...28N}. Thus the fit for carbon scaling can at best be an average. The observed scaling can be fit with the piecewise linear method as a function of [Fe/H], used for the $\alpha$-elements, well within the scatter of the data, so we use this method for simplicity. The best fit is achieved for a fiducial value of 12+log(C/H) = 8.42, rather than the B star value of 8.33.  The latter value was similar to the previous B star value reported by those authors, and somewhat lower than solar photospheric or bulk value, as discussed by \citet{2009ARAA..47..481A}. The fiducial value we propose is closer to previous solar and meteoritic values (see Table \ref{tablea1}), but  this may require revision in the light of better data and more accurate analysis methods.

The right panel shows the equivalent data for nitrogen, from \cite{2005AA...430..655S} (diamonds, halo metal-poor unmixed giants), \cite{2009AA...500.1143F} (crosses, halo solar type dwarfs and subgiants), \citet{2012AA...539A.143N} (local B stars, blue dots) and nebular data from Blue Compact Galaxies from \cite{1999ApJ...511..639I} who state that there is little evidence for dust in these objects, and, by implication, that there is little oxygen or nitrogen depletion into dust.

The determination of N is doubly difficult. Primary and secondary source behaviours that depend on individual galaxy star formation histories and particular populations, mean that the onset of secondary behaviour will vary from case to case, making a single function unlikely to match any given object. A further complication is the difficulty in estimating nitrogen abundance spectroscopically in metal poor stars, where NLTE and 3D effects need to be taken into account. Any abundance scale used in nebular modelling, based on N or N/O ratios, is somewhat uncertain. Consequently, the fit we propose here only applies to bulk well-mixed nebular abundances for the Milky Way. A check on this fit is available from the Blue Compact Dwarf galaxy nebular data in Figure \ref{fig4} from \citet{1999ApJ...511..639I} (orange discs). Although limited by the nature of the objects to a restricted metallicity range, the fit suggests that this curve provides a satisfactory description of the nebular abundance behaviour.

The stellar data for carbon and nitrogen span a range from 12+log(O/H) $\sim$6.0 to $\sim$9.0, a more extended range than is possible to achieve in nebular data, and, unlike most nebular abundances, they are not subject to dust depletion uncertainties.

 We fit a simple expression combining the primary and secondary sources to the stellar data,
\begin{equation}\label{eqn4}
\log(X/O) = \log\left[ 10^{\rm{a}} + 10^{[\log(\rm{O/H})+b]}\ \right]
\end{equation}
 where X = N or C, using $\chi$-square minimisation to obtain a best-fit analytic curves.  The fits to the stellar data are shown in Figure \ref{fig4} (red lines), with primary and secondary fits (black-dashed lines).  For carbon, a  = -0.8, b = 2.72, and for nitrogen a = -1.732, b = 2.19.  The nitrogen data show a greater scatter than for carbon, due to the intrinsic variation in nitrogen. However, the fit provides a means for scaling nitrogen abundance based on known physics. The fit of the nebular points (orange dots) suggests the curve is likely to be a considerable improvement on simple linear scaling. The observed behaviour of nitrogen was fitted analytically by \citet{2004ApJS..153....9G} for AGNs, who derived a relation similar to Equation \ref{eqn4}.

\subsubsection{Helium}
Helium was created through nucleosynthesis during the Big Bang, and is still being created continuously by stars.  Thus, the abundance of helium began at the primordial value and has increased steadily since then. In the absence of a detailed knowledge of the historical star formation rate in our galaxy, we assume a linear rate of increase with oxygen abundance. (Both oxygen and helium abundances increase due to early core-collapse supernovae, whereas the main increase in iron abundance has a delayed onset, as noted earlier). There have been numerous attempts to estimate primordial He. We adopt the primordial  abundance from WMAP measurements \citep{2004ApJ...617...29O,2008JCAP...11..012C} corresponding to a mass fraction Y$_{\rm{primordial}}$ = 0.2486 $\pm$0.0002.

Earlier pre-WMAP estimates such as that of \citet{1992MNRAS.255..325P} of 0.228 $\pm$ 0.005 could be used, but these are based on extragalactic \rion{H}{2} regions and may be affected by stellar evolutionary processes. When examining the helium content of enriched populations in globular clusters, \citet{2010MNRAS.406.1570P} adopt the value for Y$_{\rm{primordial}}$ = 0.240 $\pm$0.006 from the study of primordial nucleosynthesis by \citet{2007ARNPS..57..463S}.  It is not clear that the final value for the primordial value of the helium abundance is settled.  The WMAP measurements appear to be the most precise available and are not subject to stellar evolution modification.

Using present day helium abundance derived from the solar photosphere is unreliable, as helium has been processed by the Sun during its lifetime. Consequently we adopt the B-stars value from \citet{2012AA...539A.143N}, log(He/H) =  -1.01.   If we use the scaling factor $\zeta$ for oxygen, i.e., $\zeta_O$, we can express the abundance of helium as:
\begin{equation} \label{eqn3}
 \log(\rm{He/H}) = -1.0783+\log[ 1+0.17031\times\zeta_O/\zeta_O(0)]
\end{equation}

In the absence of better data, following \citet{1992MNRAS.255..325P}, we assume a simple linear relationship. It should be noted, however, that the same formalism can be used for any other primordial and present-day helium abundances, and if data is available to suggest a non-linear relationship with oxygen abundance, this can also be accommodated.

\subsubsection{Neon and Argon}

The noble gases neon and argon are not normally detected in the solar spectrum, so we need to seek other ways of determining how they scale. The Neon abundance has been measured in B star atmospheres  by \citet{2008AA...487..307M} and \citet{2012AA...539A.143N}, who propose values of 7.97$\pm$0.07 and 8.09$\pm$0.05 on the $12+\log$(X/H) scale, respectively. As the latter group used a revised atomic data set and applied a more rigorous analysis, we have adopted the \citet{2012AA...539A.143N} neon values for the Galactic Concordance set. We adopt the solar values for argon, derived from \cite{2015AA...573A..25S, 2015AA...573A..26S,2015AA...573A..27G}.

\begin{figure}
\centering
	\includegraphics[width=0.5\columnwidth]{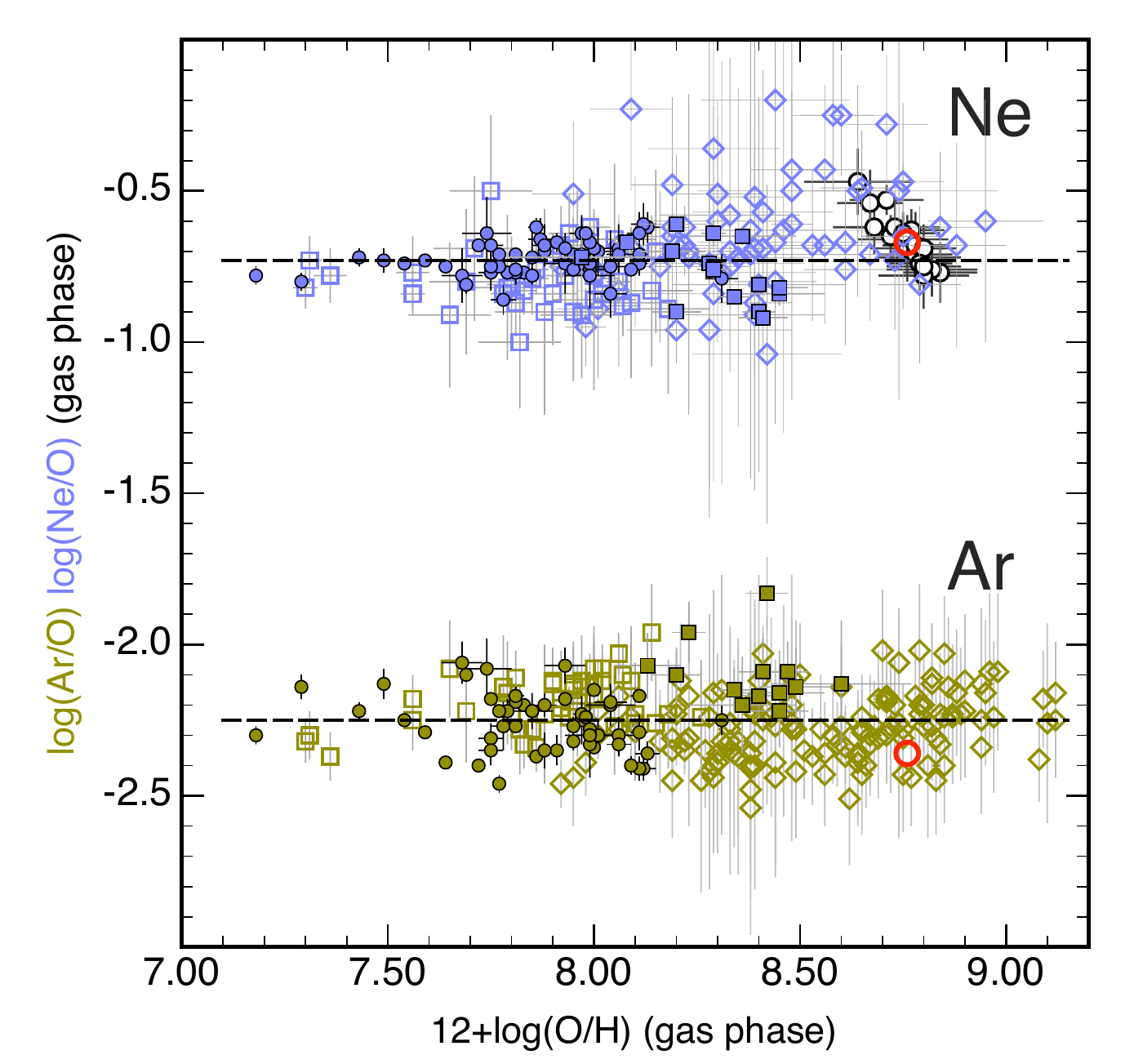}
    \caption{Scaling of Ne and Ar vs. oxygen from nebular spectra (the red circles indicate the GC fiducial values). Sources:
\citep[][squares]{1998AJ....116.2805V},
\citep[][circles]{1999ApJ...511..639I},
\citep[][diamonds]{2006ApJ...636..214V},
\citep[][filled squares]{2013ApJ...775..128B}
}
    \label{fig5}
\end{figure}

Both neon and argon are readily detected in the spectra of \rion{H}{2} regions, and as $\alpha$-elements, we can use oxygen scaling with iron as a guide. Figure \ref{fig5} presents plots of log(Ne/O) and log(Ar/O) vs 12+log(O/H) taken from several nebular sources. The red circles indicate the galactic concordance fiducial values. Note that these indicate the total abundances, whereas the nebular data only record the gas phase abundances, making no allowance for the dust depletion, which is likely to be variable and is not well known.  It is unlikely that any neon or argon is depleted into dust, as indicated by the very low abundances in chondritic meteorites \citep{2003ApJ...591.1220L, 2009LanB...4B...44L}. In Figure \ref{fig5}, error bars have been included to illustrate that constant linear fits are warranted in both cases.  Because the nebular data do not allow for dust depletion, the total (dust and gas phase) oxygen is greater than the plotted nebular points indicate, and one would expect the GC origin points to be below the linear fits. The GC fiducial value for neon from the B star data \citep{2012AA...539A.143N} is above the fit line, whereas dust depletion of oxygen would suggest it should be below. The sizes of the error bars do not allow us to draw any conclusion, so for consistency we have retained the B star value. For argon we have adopted the most recent solar value (see Table \ref{tablea1}). Both values may need to be adjusted in the light of planned modelling of particular \rion{H}{2} regions in the Magellanic Clouds, the subject of forthcoming papers.

\subsubsection{Chlorine}

Solar photospheric abundances of chlorine can only be measured (indirectly) in sunspots from the HCl abundances \citep[][and references therein]{2009ARAA..47..481A}. The solar values for chlorine from \cite{2009ARAA..47..481A} date back with minor variations to sunspot measurements, ca. 1970.

\begin{figure}
\centering
	\includegraphics[width=0.5\columnwidth]{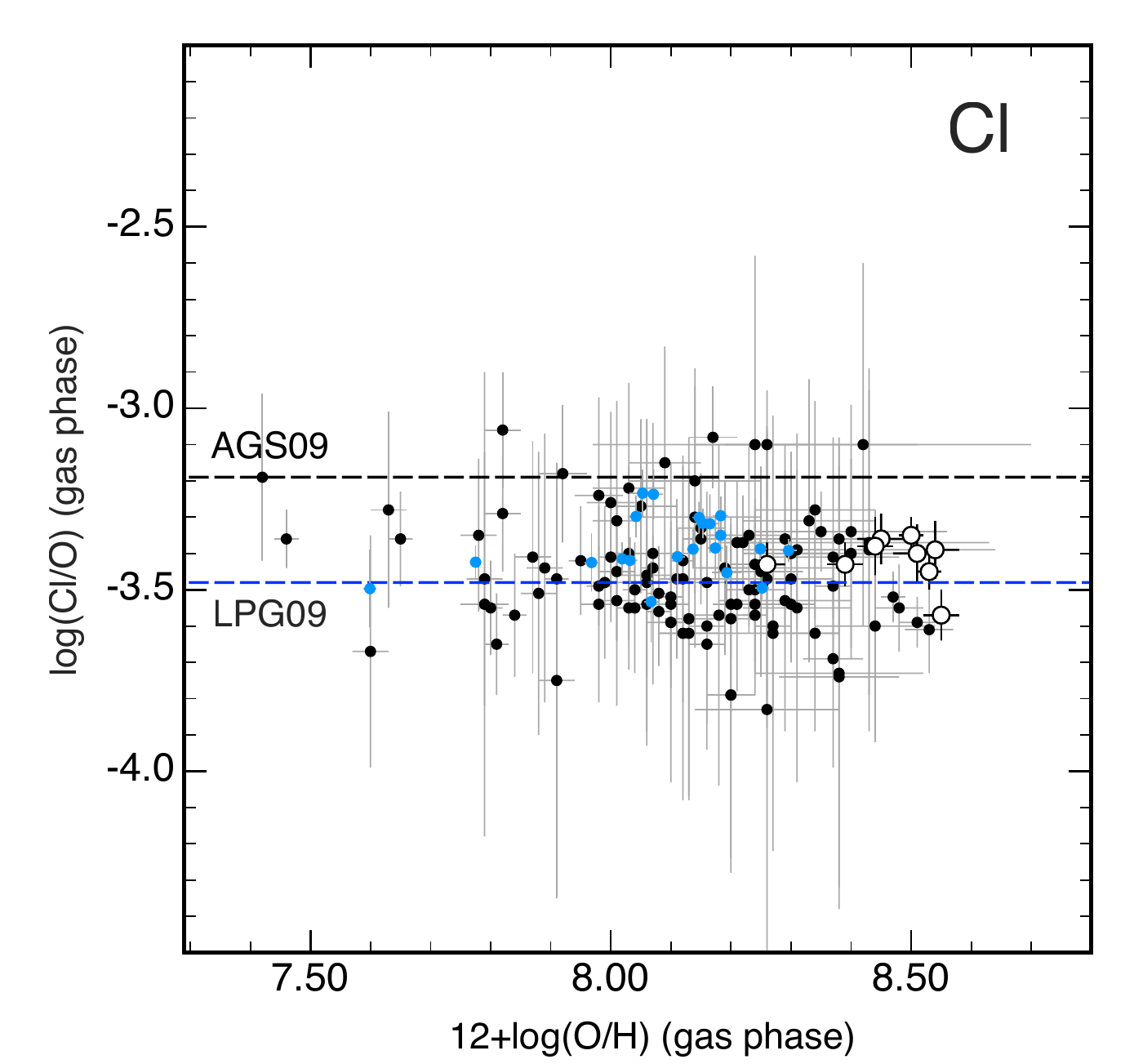}
    \caption{Nebular chlorine vs oxygen: the dashed lines are from \citet{2009ARAA..47..481A} (AGS09, black line, solar) and \citet{2009LanB...4B...44L} (LPG09, blue line, meteoritic). The circles are gas-phase Milky Way nebular data from \citet{2015MNRAS.452.1553E}. The dots are from gas-phase nebular data from \citet{2004ApJ...602..200I} (blue, Blue Compact Galaxies) and \citet{2006AA...448..955I} (black, metal poor emission line galaxies).}
    \label{fig6}
\end{figure}

However, gas-phase abundances of chlorine have been measured from high resolution nebular spectra \citep[e.g.,][]{2007ApJ...670..457G}. Recently, \cite{2015MNRAS.452.1553E} presented revised values for nebular chlorine as it scales with oxygen in Milky Way \rion{H}{2} regions. Figure \ref{fig6} shows log(Cl/O) vs 12+log(O/H) from Milky Way \rion{H}{2} regions \citep{2015MNRAS.452.1553E} and from extra-galactic \rion{H}{2} regions \citep{2004ApJ...602..200I, 2006AA...448..955I}. The horizontal dashed lines show the scaling for solar photosphere data from \citet[][AGS09, black dashed line]{2009ARAA..47..481A} and meteoritic data from \citet[][LPG09, blue dashed line]{2009LanB...4B...44L}, assuming chlorine scales with oxygen. The nebular abundances have not been corrected to total abundance (gas-phase plus dust), because dust depletion for oxygen and chlorine are variable and not well known.

Chlorine is likely to be depleted into moderately volatile compounds \citep{2003ApJ...591.1220L}, and can react efficiently with neutral hydrogen to form H+Cl compounds \citep{2015AA...575L...8B, 2012ApJ...744..174M}, so it is likely that it will be somewhat depleted in \rion{H}{2} regions. Thus the nebular data cannot be used to define a generic value for chlorine abundance, but they do suggest a better fit to the meteoritic data than the solar photospheric data (Figure \ref{fig6}), although \cite{2003ApJ...591.1220L} notes that chlorine in meteorites is variable. For these reasons, in the Galactic Concordance we adopt the meteoritic values for chlorine from \cite{2009LanB...4B...44L}.

Although they are not prominent, emission lines of chlorine [\rion{Cl}{2}] and [\rion{Cl}{3}] are observed in low noise \rion{H}{2} region spectra. Chlorine does not play a significant role in the thermal balance of nebulae, but chlorine lines, where observed in nebulae, can be a useful density diagnostic and the reference abundance may warrant revision when better data is available.

\subsubsection{Super-solar metallicity stars}

Some stellar data is available for Z $>$ 1.0, for example, \citet{2011AA...535A..42T} and the massive open cluster NGC6791 \citep{2012ApJ...756L..40G}. However, some of the data are from old, very metal-rich galactic disk stars, which have undergone substantial evolution and exhibit photospheric enrichment. The data do not provide a useful model for abundance scaling, but do suggest, however, that the second break point in the piece-wise linear fit might be at a value of [Fe/H] $>$ 0.5, so this parameter may need revising when data for metal-rich main sequence stars becomes available.

Some elements (e.g., Na, Sc, V, Co and Ni, see Appendix A) appear to exhibit a slight upturn for [Fe/H] $>$ 0.  We have not attempted to model this behaviour for two reasons. Most important is that none of these elements plays a major role in nebular thermal balances. Second, if the reality of the upturn is confirmed with more extensive data, it will be possible to accommodate it using the $\Delta$ parameter described by Equation \ref{eqn8}, below.

\subsection{Minor nebular elements}

Apart from the 12 elements considered above, there are others that are of minor importance in the energy balance in \rion{H}{2} regions and other emission nebulae, but may be important in modelling stellar atmospheres and evolutionary tracks. The adopted reference abundances for the remaining 18 elements to Zn and their sources are given in Table \ref{table1}. The stellar data and scaling fits for these elements are in Appendix A.

\section{The nebular scaling function}

\subsection{General approach}

The extensive stellar data demonstrate that different populations of stars in the Milky Way (and other galaxies)  are present that differ somewhat in their scaling behaviour with [Fe].  The objective of this work is to propose a  series of linear fits to the bulk MW stellar trends as a first order approximation, rather than to attempt different fits for specific populations.  The latter is possible, but for the primary purpose of this work. Piece-wise linear fits to the bulk trends are appropriate to describe the scaling relations.

\subsection{Simple fits to the observed stellar abundance scaling}

To model this behaviour, we adopt simple piece-wise linear fits, as shown by the red lines in the stellar data plots. For the bulk trends, these need only be fit by eye, as a $\chi$-square or least-squares fit is not warranted for our purposes, and the statistical variability of the data points may not be Gaussian, rendering such fits inappropriate. However, Figure \ref{fig2} (left panel) shows a least squares fit, which is close to the adopted fit and well within the scatter of the data. While the real behaviour could exhibit curved breakpoints, the intrinsic scatter in the data (and the measurement uncertainties, not shown for clarity) do not warrant a more complex model. This model has the additional benefit of computational simplicity. The standard (GC) metallicity (which we refer to subsequently as the ``fiducial point'') is marked with a yellow circle in Figure \ref{fig2} and subsequent plots. The piecewise linear fit for oxygen may be expressed as:
{\small
\begin{eqnarray}\label{eqn5}
[\rm{O/Fe}] & = & +0.50  \; ,\; -2.5 < [\rm{Fe/H}] < -1.0 \;,\nonumber \\
            & = & -0.5\times[\rm{Fe/H}] \;,\; -1.0 < [\rm{Fe/H}] < 0.5 \;,\nonumber \\
            & = & -0.25 \; , \; [\rm{Fe/H}] > 0.5 \; .
\end{eqnarray}
}

The initial 0.5 factor is characteristic of the initial O/Fe yield in massive early stars.  Similar factors apply to other $\alpha$-elements, but with different values. We call this factor $\Xi$ and as the scaling is relative to iron, we append the Fe suffix: $\Xi_{Fe}$. For oxygen on the iron scale the factor is $\Xi_{Fe}(\rm{O})$, for magnesium, $\Xi_{Fe}(\rm{Mg})$, etc.

As the $\alpha$-elements, to a good approximation, share the same break points, Equation \ref{eqn5} can be generalised to describe  iron-based scaling for any element X with the same breakpoints:
{\small
\begin{eqnarray}\label{eqn6}
[\rm{X/Fe}] & = &+\Xi_{Fe}(\rm{X}) \;,\; -2.5 < [\rm{Fe/H}] < -1.0 \;, \nonumber \\
            & = &-\Xi_{Fe}(\rm{X}) \times[\rm{Fe/H}]  \;,\; -1.0 < [\rm{Fe/H}] < 0.5 \;, \nonumber\\
            & = &-\Xi_{Fe}(\rm{X}) \times 0.5  \;,\; [\rm{Fe/H}] > 0.5 \;.
\end{eqnarray}
}

Below [Fe/H] =-2.5 the stellar data are too sparse to warrant a fit. Our aim here is to establish general fits as the basis for improved abundance scaling in photoionisation models, recognising that the detailed abundance behaviour may be somewhat more complex, and/or variable. The intrinsic spread of the data in all the graphs provides an estimate of the errors in the $\Xi$ parameters derived for each element, approximately $\pm$20\%.

\subsection{Separating the components}

In section 2.4 we introduced the scaling parameter $\zeta$ (referred to the chosen scaling base element, usually iron or oxygen), and the Galactic Concordance reference abundance set.  To describe the scaling of individual elements, taking into account their different scaling behaviours, we use a general expression to separate the specific behaviour of each element from the fiducial value and the scaling parameter.

We introduce the parameter $\Delta$ (dex) to describe individual element behaviours. $\Delta$ encompasses the evolutionary details, i.e., the way the abundance of an element scales with $\zeta$ in the scaling base chosen, for example, Fe:
\begin{equation}\label{eqn7}
     \log(\rm{X/H}) = \log(\rm{X/H})_0 + \Delta_{Fe}(X) + \log \zeta_{Fe}
\end{equation}
where the zero suffix refers to the fiducial value for the element X. For simple scaling, $\Delta$ = 0. For piecewise linear iron-base scaling, such as the $\alpha$-elements exhibit (Equation \ref{eqn6}), with a low abundance ratio $\Xi_{Fe}$ and abundance break points $\chi_0$ and $\chi_1$ (dex),  $\Delta_{Fe}$ for a given element X is a function of $\zeta_{Fe}$, $\Xi_{Fe}$, $\chi_0$, and $\chi_1$:
{\small
\begin{eqnarray}\label{eqn8}
\Delta_{Fe}(\rm{X}, \zeta_{Fe}) & = &\Xi_{Fe}(X) = \rm{const.} \; , \;  \log \zeta_{Fe} < \chi_0 \;, \nonumber \\
          & = &\frac{\Xi_{Fe}(\rm{X})}{\chi_0} \times \log \zeta_{Fe} \; , \;  \chi_0 < \log \zeta_{Fe} < \chi_1 \;,\nonumber \\
          & = &\frac{\Xi_{Fe}(\rm{X})}{\chi_0} \times \chi_1 = \rm{const.} \; , \;  \log \zeta_{Fe} > \chi_1\ .
\end{eqnarray}
}

The second part of the fit passes through [Fe/H] = 0 at log($\zeta_{Fe}$) = 0. While it would be possible to fit a more complex function, given the uncertainties in the data, a high order function for $\Delta$ is not warranted.

However, nitrogen is not well described by a simple piecewise linear fit, because of the complexities of primary and secondary enrichment (Figure \ref{fig4}, right panel, and Equation \ref{eqn3}). This case illustrates how $\Delta$ can be generalised to more complex forms, using oxygen as the scaling base:
\begin{equation}\label{eqn9}
      \Delta_O(\rm{N}, \zeta_O) = \log\left( 10^{-0.764} + 10^{\log \zeta_O - 0.082} \right)
\end{equation}

Equation \ref{eqn8} demonstrates an important aspect of abundance scaling. As Fe scales, for example, at low Fe/H values, oxygen is enhanced considerably:
\begin{equation}\label{eqn10}
     \log(\rm{O/H}) = \log(\rm{O/H})_{\rm{O}} + \Delta(\rm{O},\zeta_{\rm{Fe}}) + \log \zeta_{\rm{Fe}}
\end{equation}
where $\zeta_{\rm{Fe}}$ is linear in Fe enrichment and $\Delta = \Xi_{\rm{Fe}}(\rm{O})$, or $\sim$~+0.5 dex at low Fe/H. In other words, the well known enhancement of oxygen relative to iron at low metallicity is expressed explicitly.

\subsection{Changing scales}

Using this system it is easy to convert from one $\zeta$ scale to another, for example, to use oxygen scaling, allowing iron to be depleted at low oxygen enrichments, we have:
\begin{equation}\label{eqn11}
      \log \zeta_O = \Delta(O, \zeta_{\rm{Fe}}) + \log \zeta_{\rm{Fe}}
\end{equation}
where $\Delta$ is the same function used for Fe scaling. The break points $\chi_{0,1}$ have different values on the $\zeta_O$ and $\zeta_{Fe}$ scales, but can be converted simply from one scale to the other.

If a different scaling for element Y is required, the $\Xi_X$ of all elements (Z) are converted to $_Y$ simply via:
\begin{equation}\label{eqn12}
      \Xi_Y(Z)  = \Xi_X(Z) - \Xi_X(Y)
\end{equation}
Note $\Xi(X)_X \equiv 0$, so $\Xi(X)_Y = -\Xi(Y)_X$
and the break points (in dex) by:
\begin{equation}\label{eqn13}
      \chi_0(Y) = \chi_0(X) + \Xi_X(Y)
\end{equation}
and
\begin{equation}\label{eqn14}
      \chi_1(Y) = \chi_0(Y) \times \chi_1(X)/\chi_0(Y)
\end{equation}
and then the $\zeta_Y$ and $\Delta_Y$ parameters for the new element are available.

\subsection{Graphical illustration}

\begin{figure}
\centering
	\includegraphics[width=0.7\columnwidth]{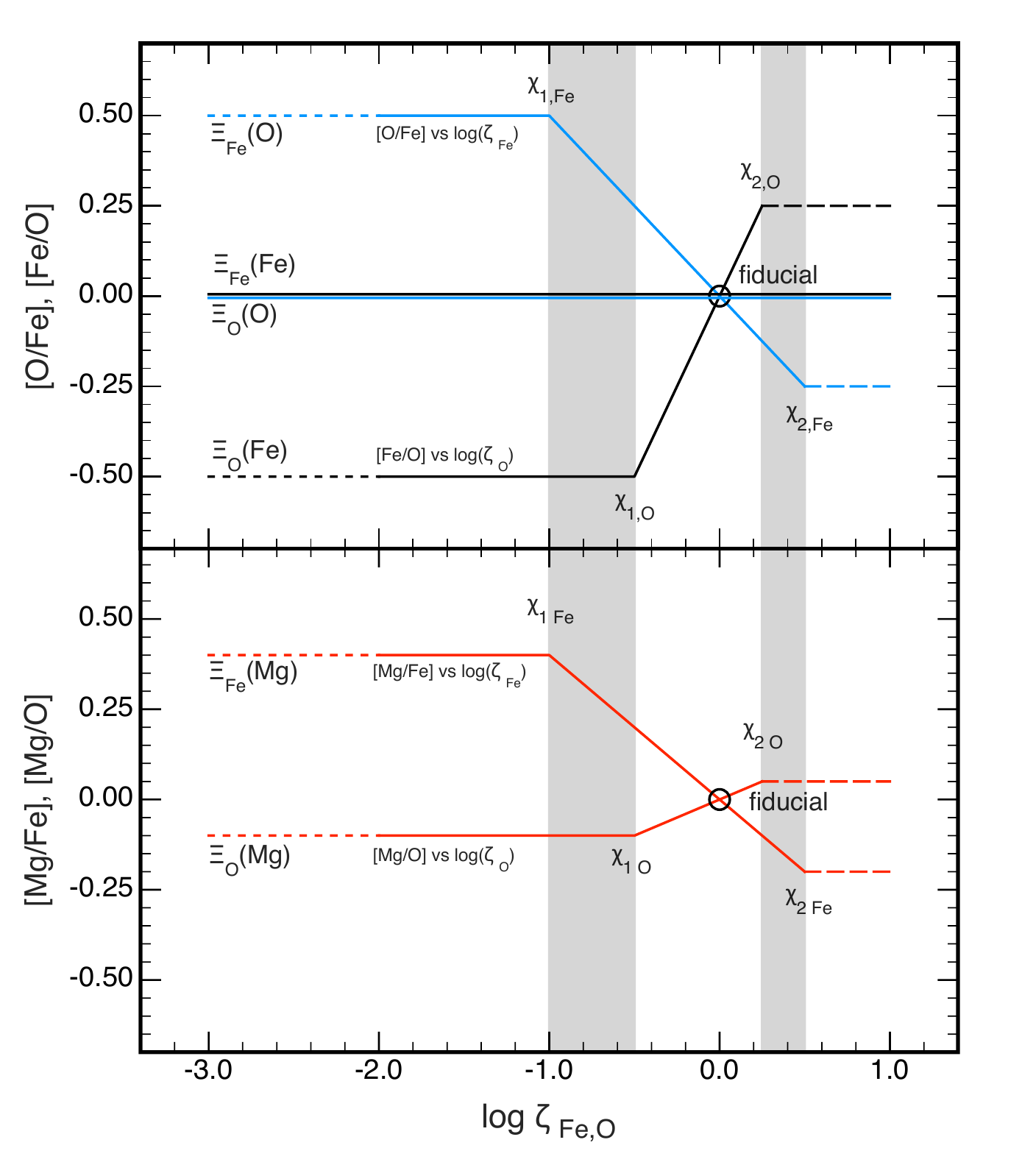}
    \caption{Upper panel: Plots of [X/Fe] vs log$\zeta_{\rm{Fe}}$ (blue lines, X = Fe and O) and [X/O] vs $\log \zeta_{\rm{O}}$ (black lines, X = Fe and O). Lower panel: Plots of [Mg/Fe] vs $\log \zeta_{\rm{Fe}}$ and [Mg/O] vs $\log \zeta_{\rm{O}}$
    }
    \label{fig7}
\end{figure}

Figure \ref{fig7} illustrates equations \ref{eqn7}-\ref{eqn14} graphically. The upper panel shows [X/Fe] vs $\log \zeta_{\rm{Fe}}$ (blue lines, where X = Fe and O) and [X/O] vs $\log \zeta_{\rm{O}}$ (black lines, where X = Fe and O). The shorter dashes indicate where the fit is not reliably based on stellar data, the longer dashes are the assumed behaviour, without stellar evidence. The lower panel presents similar data for magnesium, showing how the scaling changes when the scale base element is changed. Note that for the x-axes,
\begin{equation}\label{eqn15}
\log \zeta_{\rm{O}} \equiv \log(O/H) - \log(O/H)_{\rm{fiducial}} \equiv [O/H]
\end{equation}
and
\begin{equation}\label{eqn16}
\log \zeta_{\rm{Fe}} \equiv \log(Fe/H) - \log(Fe/H)_{\rm{fiducial}} \equiv [Fe/H].
\end{equation}
The grey shaded areas illustrate the different values of the break points between Fe and O.

\begin{figure}
\centering
	\includegraphics[width=0.7\columnwidth]{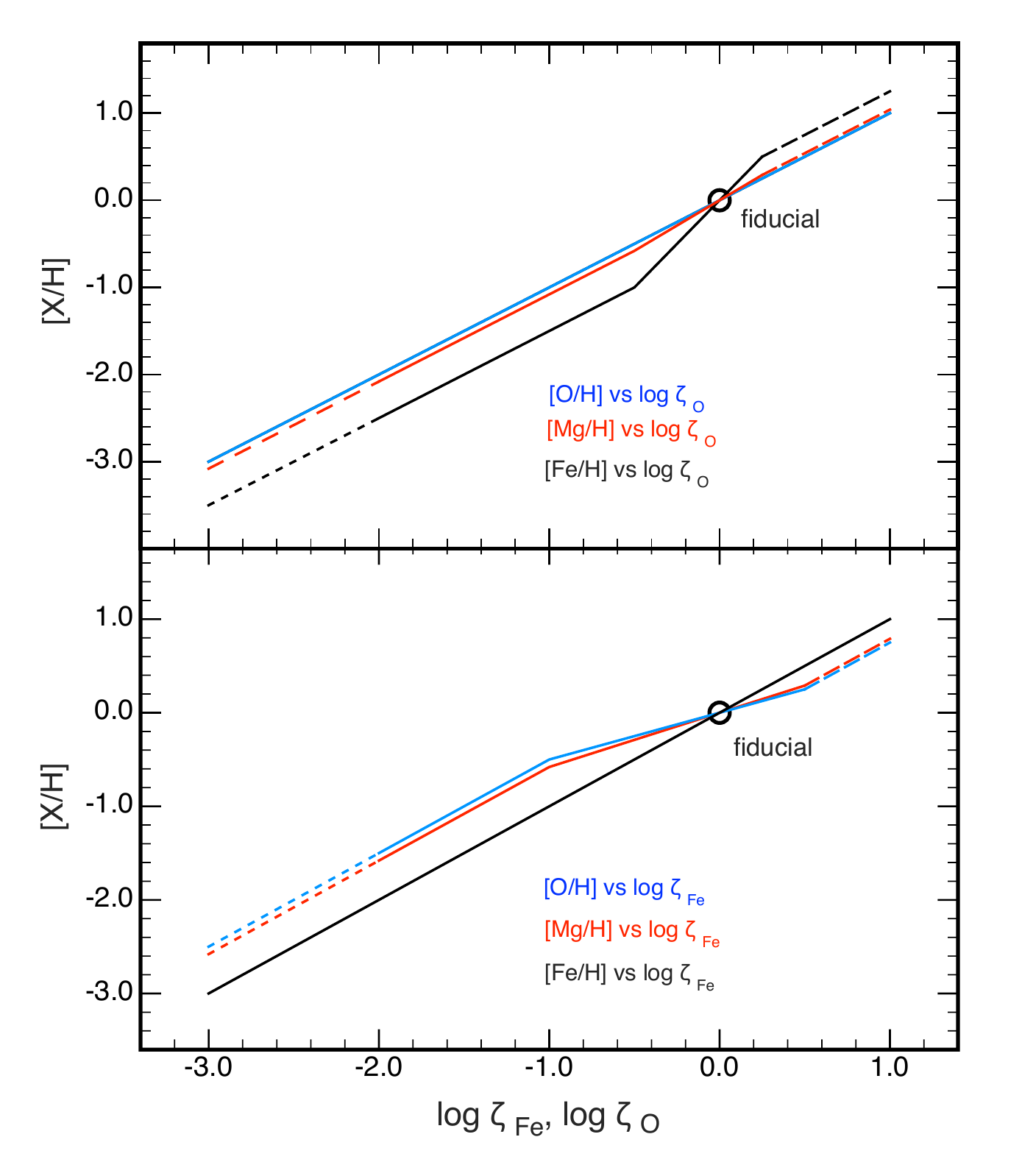}
    \caption{Upper panel: Plots of [X/H] vs log$\zeta_O$ for iron, oxygen and magnesium (X). Lower panel: vs $\log \zeta_{\rm{Fe}}$}
    \label{fig8}
\end{figure}

Figure \ref{fig8} shows plots of [X/H] vs log$(\zeta_{\rm{Fe}})$ for iron, oxygen and magnesium (X) (upper panel), and [X/H] vs log($\zeta_{\rm{Fe}}$). The behaviour of the different scalings is much easier to understand in Figure \ref{fig7}, and is why we have used this approach in Figures \ref{fig2}-\ref{figA4} and \ref{fig9}.

\subsection{Summary of scaling parameters}

Table \ref{table2} summarises the scaling parameters for Milky Way stellar abundances for hydrogen to zinc, with the low metallicity level $\Xi$ and the break points $\chi_0$ and $\chi_1$ expressed in the Fe base scale, and also converted to the oxygen base scale. The fiducial scale $\log(X/H)_0$ is included from Table \ref{table1}, less the 12 factor. For the $\alpha$ and $\alpha$-like elements, $\Xi_{Fe}$ and the two break points are derived from the observed stellar abundances. For iron-peak and iron-peak-like elements, $\Xi_{Fe}(\rm{X})$ = 0. Because of the scatter on the stellar values and differences between stellar populations, precise fitting of the model is not possible.

\begin{table}
\small
\centering
\caption{Scaling parameters $\Xi_{\rm{Fe}}$ and $\Xi_{\rm{O}}$ for Milky Way stars, with upper and lower break points. For nitrogen the behaviour is not well described by the piecewise linear fit model, but a value can be ascribed to $\Xi$ and the primary/secondary scaling specified by the $\Delta$ parameter (Equations \ref{eqn8} and \ref{eqn9}). Carbon can be sufficiently well described by the piecewise linear fit and a single $\Xi$ value. The elements H, He, Li, Be and B are not described by the $\Xi$ parameter, because hydrogen is the reference element, we assume helium scales simply with $\zeta_O$ and Li - B are not important in nebular analysis, as well as having very low abundance. For convenience the 'fiducial' value of log(X/H) is repeated from Table 1, without the  addition of 12.}
\label{table2}
\begin{tabular}{llllr} 
\hline
Z & X & $\Xi_{\rm{Fe}}(X)$ & $\Xi_{\rm{O}}(X)$ & log(X/H)$_0$\\
\hline
 & \textit{upper break} & -1.0 & -0.5 & \\
 & \textit{lower break} & 0.50 & 0.25 & \\
\hline
1 & H & - & - & 0.000 \\
2 & He & - & - & -1.010 \\
3 & Li & - & - & -8.722 \\ 
4 & Be & - & - & -10.680 \\
5 & B & - & - & -9.193 \\
6 & C & 0.063 & -0.437 & -3.577 \\
7 & N & (-0.264) & (-0.764) & -4.210 \\
8 & O & 0.500 & 0.000 & -3.240 \\
9 & F & 0.500 & 0.000 & -7.560 \\
10 & Ne & 0.500 & 0.000 & -3.910 \\
11 & Na & 0.200 & -0.300 & -5.790 \\
12 & Mg & 0.400 & -0.100 & -4.440 \\
13 & Al & 0.400 & -0.100 & -5.570 \\
14 & Si & 0.400 & -0.100 & -4.500 \\
15 & P & 0.000 & -0.500 & -6.590 \\
16 & S & 0.400 & -0.100 & -4.880 \\
17 & Cl & 0.500 & 0.000 & -6.750 \\
18 & Ar & 0.500 & 0.000 & -5.600 \\
19 & K & 0.400 & -0.100 & -6.960 \\
20 & Ca & 0.350 & -0.150 & -5.680 \\
21 & Sc & 0.250 & -0.250 & -8.840 \\
22 & Ti & 0.350 & -0.150 & -7.070 \\
23 & V & 0.000 & -0.500 & -8.110 \\
24 & Cr & 0.000 & -0.500 & -6.380 \\
25 & Mn & 0.000 & -0.500 & -6.580 \\
26 & Fe & 0.000 & -0.500 & -4.480 \\
27 & Co & 0.000 & -0.500 & -7.070 \\
28 & Ni & 0.000 & -0.500 & -5.800 \\
29 & Cu & 0.000 & -0.500 & -7.820 \\
30 & Zn & 0.200 & -0.300 & -7.440 \\ \hline

\end{tabular}
\end{table}

For F, Cl, Ne and Ar, there are no extensive stellar abundance scaling data and we assume their abundances scale with oxygen, so the $\Xi_{Fe}$ values are that of oxygen. The abundance data for Li, Be, B are taken from meteoritic values and we have little information on how they scale. As they do not play a major role in nebular physics, they can safely be ignored here. Carbon can be scaled equally well to iron with a low value of $\Xi_{Fe}$ or to the primary/secondary curve fit, given the spread in the data, so the former was chosen for simplicity of computation.  For nitrogen we have used the primary/secondary fit curve, scaled with oxygen.

The data in Table \ref{table2} are intended as a general guide to how typical Milky Way thick disk stellar abundances scale.  Precision is not possible from the available data, and it is likely that for any element, no one single value of $\Xi$ applies to all stellar populations sampled. As noted, carbon also has a primary/secondary growth curve (Figure \ref{fig4}, left panel), but given the observational uncertainties, abundances can be sufficiently well described by the piecewise linear fit and a single $\Xi$ value. We use the scaling described by these ``average'' $\Xi$ values as the basis for nebular scaling in the Milky Way and similar galaxies, to replace the simple scaling previously assumed.

\subsection{Web abundance calculator}

An online web application has been implemented that allows all of the above scaling calculations to be computed: \url{http://miocene.anu.edu.au/mappings/abund}. This paper provides the background necessary to use that application.

\subsection{Scaling in other galaxies}

\begin{figure}
\centering
	\includegraphics[width=0.5\columnwidth]{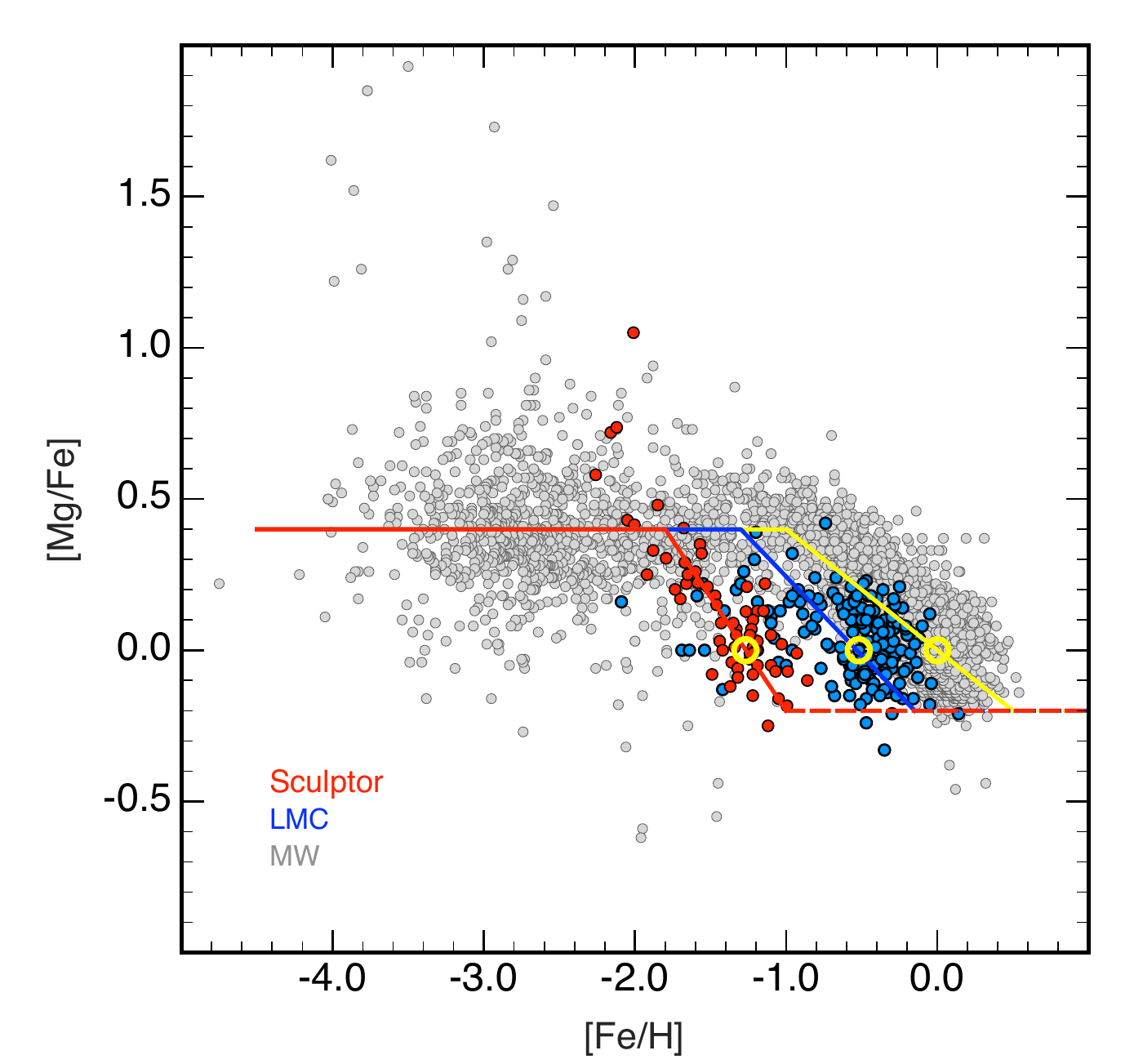}
    \caption{[Mg/Fe] vs [Fe/H] for Milky Way (grey circles), LMC (blue circles) and the Sculptor Dwarf (red circles) \citep[MW:][]{2010AN....331..474F, 2011ApJ...737....9R, 2012AA...545A..32A, 2013AA...552A...6G, 2014AA...562A..71B, 2014AJ....148...54H}, \citep[LMC:][]{2008AA...480..379P, 2013AA...560A..44V}, \citep[Scl dwarf:][]{2005AJ....129.1428G, 2009ApJ...705..328K, 2010AA...524A..58T}.  A common value for the low-metallicity trend for the three objects is expected for an invariant massive star IMF \citep{1998ASPC..142...89W}. The different break points and gradients in the three galaxies reflect their different rates of specific star formation and enrichment. The fiducial point circles also differ for each object, all marked in yellow for visibility.}
    \label{fig9}
\end{figure}

All the stellar data considered so far have been derived from Milky Way stars. The spectra of individual giant stars can also be measured for nearby galaxies, for example the Large Magellanic Cloud (LMC) \citep{2008AA...480..379P, 2013AA...560A..44V} and the Sculptor Dwarf Elliptical galaxy \citep{2010AA...524A..58T,  2005AJ....129.1428G, 2009ApJ...705..328K}. Figure \ref{fig9} shows the same data for [Mg/Fe] vs [Fe/H] as in Figure \ref{fig2} but including data for the Sculptor Dwarf and the LMC bar and disk. The figure shows the Sculptor Dwarf (red circles), the LMC (blue circles) and the Milky Way (grey circles). The Sculptor stars exhibit same behaviour as the Milky Way stars, but with a lower [Fe/H] breakpoint and steeper drop.  This is to be expected, as there would have been fewer core-collapse supernovae in the Sculptor Dwarf before the Type 1a supernovae started enriching iron, compared to the Milky Way. The LMC stars also exhibit a lower break point than the Milky Way, due to lesser contributions to the metallicity from core-collapse supernovae, as suggested by  \cite{2013AA...560A..44V}. The LMC appears to have two populations as some stars have abundance ratios similar to the Milky Way stars, while others appear to be intermediate between the Milky Way and Sculptor. The piecewise linear fit lines are chosen by eye to illustrate the trends.  The scaling behaviour in each galaxy is similar in form, suggesting a universal process, but one where the break point and ``zero point'' depend on the galaxy, as described by \citet{1993ASPC...48..727W}.  Stars in the Fornax Dwarf behave similarly to the Sculptor Dwarf \citep{2010AA...524A..58T}, but have been omitted for clarity.

We can draw an important conclusion from this. While the piece-wise linear scaling of $\alpha$-elements with Fe appears to be a universal process, the break point depends on the star formation history \citep{1993ASPC...48..727W} and thus on the galaxy mass. Where the star formation history of larger galaxies is similar to that of the Milky Way, the abundance scaling for these galaxies can be assumed to follow the scaling fits derived for the Milky Way. However, \citet{2015MNRAS.450..342K} found three distinct groups of star forming galaxies in the local region. Modelling these groups may require modification to the break points, but this must await observations of appropriate stellar populations and their abundances. It is possible that careful fitting of the abundance behaviour of the different populations in the Milky Way may cast light on the variability of the fit parameters in other galaxies. Modelling the abundance behaviour of small dwarf galaxies is also likely to require different scaling than for the larger spirals. Calibrating the break point to star formation history might provide useful information.

The data in Figure \ref{fig9} do not allow us to decide whether there is a common value for the low-metallicity trend for the three objects in Figure \ref{fig9}. As the early enrichment is due to core-collapse supernovae, and it is possible that the massive star IMF is invariant \citep{1998ASPC..142...89W,2015AA...582A.122K}, once we have a statistically mixed ISM, the [Mg/Fe] to [Fe/H] ratio should be consistent between different galaxies at low metallicities. Clearly, more data is required to confirm this. In general, however, the scaling behaviour follows a similar pattern. Appropriate fits to the parameters that specify the trends (fiducial point, break point and low metallicity trend level) will most probably describe the scaling in any galaxy. Better observational data is required to identify appropriate parameters.

\section{Dust depletion in \rion{H}{2} regions}

The quantity and composition of dust in \rion{H}{2} regions is an ongoing question in nebular physics. For example, \citet{1993MNRAS.261..306H} found that large errors could be introduced into C, N and O abundance estimates in \rion{H}{2} regions due to dust depletion.  It appears likely that the dust amount increases with total  metallicity, due to the greater availability of refractory elements \citep[see, for example,][Figure 1]{2011AA...532A..56G}. Both the amount and composition of dust may vary significantly, due to inhomogeneities in the distribution of dust-forming elements in supernova remnants and AGB star enrichment embedded in \rion{H}{2} regions. This problem is likely to be especially complex in measuring the abundances in active star forming regions and in high redshift galaxies, as discussed  in \citet{2003Natur.423...57P}.

There are a number of approaches that can be taken to estimate the level of dust depletion. First, direct measurements of absorption lines in the spectra of nearby stars can be used to measure the abundance of dust, e.g.,  \citet{2009ApJ...700.1299J}, who summarised and modelled a range of elements, and \citet{1996ARAA..34..279S} and, \citet{1996ApJ...457..211S}, who investigated Milky Way gas and dust abundances in the warm neutral medium, in cold diffuse clouds, and in distant halo clouds. The nature of the dust in the ISM  is likely to be different than in giant molecular clouds, and in ionised regions of these, but ISM measurements may be useful as a starting point for modelling depletions.

Second, as explored by \citet{1998MNRAS.295..401E, 2004MNRAS.355..229E}, comparing the observed abundances, especially of O, Mg, Si and Fe, in the atmospheres of young stars associated with \rion{H}{2} regions, if they can be measured, with the observed abundances from nebular emission lines can give a useful measure of the depletion levels.

In a third approach, starting with some degree of a priori knowledge of depletion ratios and amounts, we can explore the emission regions with photoionisation models, to see how well different dust depletion assumptions match the observations. To explore this idea, we have obtained high resolution, low noise integral field spectra of several \rion{H}{2} regions in the Milky Way, LMC and SMC where the geometries are simple and therefore can be modelled with some accuracy, as the basis for comparing with models.

None of these methods is ideal, so educated combinations of such methods may need to be used. The abundance scaling relations presented here suggest a starting point for estimating dust depletions, based on the availability of elements to form dust and augmented by estimates of relative depletions. Where data is available by comparison of central star cluster stellar abundances and apparent nebular abundances, this can provide an independent check on the accuracy of the modelling method.

A detailed analysis of the effects of dust depletion is beyond the scope of the present work. At this point it is reasonable to suggest that more accurate scaling of abundances is likely to provide a better basis for estimating dust depletions, providing better estimates of the total abundances in nebulae.

\section{Implications for stellar population synthesis models}

Stellar population synthesis models are an important component in the photoionisation modelling of \rion{H}{2} regions, as they provide a guide to the ionising radiation energising the nebulae. Perhaps the most widely used in nebular modelling is $\textsc{Starburst99}$ \citep{1999ApJS..123....3L,2005ApJ...621..695V,2012ApJ...751...67L,2014ApJS..212...14L}. This takes as its input model spectra of stellar atmospheres and stellar evolutionary tracks for a variety of stellar masses and metallicities.  These models have evolved over the past 30 years, and the population synthesis applications have adapted to the new inputs \citep{2014mysc.conf..157L}. However, the atmosphere and evolutionary track models are not fully compatible, due in part because they were created for different purposes. In particular, there is a problem with the standard abundances used. For example, the stellar atmosphere models from \citet{2001AA...375..161P} and \citet{2003ApJ...599.1333S}  use the solar abundance values from \citet{1989GeCoA..53..197A} or earlier, and iron-based scaling. The evolutionary track models \citep[][and references to older models therein]{2012AA...537A.146E, 2013AA...553A..24G, 2014AA...566A..21G} are based on \citet{1989GeCoA..53..197A} and \citet{2009ARAA..47..481A} solar values, and use metallicities that are oxygen scaled.  Consequently, matching tracks to atmospheres to generate a synthetic stellar population is difficult. As far as possible, when modelling nebular emission, consistent abundances between the tracks, atmospheres and the total nebular metallicities should be used. This restricts the metallicities available for modelling to values  that have some basis in the input stellar models.  The ability to ``translate'' between the different solar scales and scaling base elements provided in this paper give us some confidence that we are matching nebular models to stellar atmospheres and evolutionary tracks as best we can.

To reduce the problems arising from the shortcomings if current stellar atmosphere models, we are developing a new grid of models of hot stars as inputs to population synthesis applications, using the CMFGEN code \citep{2012IAUS..282..229H}. We are also working with another group to develop evolutionary track models using metallicity scales consistent with the stellar atmosphere models, using the MESA code \citep{2016ApJ...823..102C}.  This should allow us to avoid the current scale discrepancies in population synthesis calculations.

\section{Commentary and Conclusion}
The problem of ongoing changes in ``solar'' abundance values is a source of uncertainty in photoionisation models. It is conventional in detailed physical models (Cloudy, $\textsc{Mappings}$, etc) to use a set of stellar atmosphere models, sets of stellar evolutionary tracks, combined into a model excitation spectrum to ionise the nebula using population synthesis applications.

Even ignoring the deficiency of hot star models in the stellar atmosphere data, there is a lack of consistency between the solar metallicity scales used in stellar atmospheres and the evolutionary tracks. Fixing this inconsistency by developing stellar atmosphere models and evolutionary tracks using the same reference abundances and scaling behaviour will be an important step in improving the reliability of nebular models.

To avoid these problems, we advocate a reference scale derived from a population of present day B stars as a sensible standard. \citet{2012AA...539A.143N} provide such a basis, for all the elements that play an important part in nebular physics. We have augmented the scale with elements of lesser important from other scales, including  the latest solar and the most carefully measured meteoritic scales.  However, from the perspective of nebular modelling, this is far less important than setting a reliable scale for the key elements.

We present a new scaling parameter, $\zeta$, to specify the  metallicity, and to replace the usual Z/Z$_{\odot}$ parameter.  We derive new scaling relations from the stellar spectra to describe the way in which elemental abundances scales in ionised nebulae. A consequence of these relations is a simple method to convert between different elements as the base for scaling measurement---usually oxygen for nebular analysis and iron for stellar analysis. We provide a simple web-based application to allow easy calculation and conversion of abundances, based on this paper, available at \url{http://miocene.anu.edu.au/mappings/abund}.

We are not suggesting the Galactic Concordance scale is the only one that can be used, just that it is important to use a single scale for all the components involved in photoionisation models. The scale conversion methods we propose can be used to convert between any scale the photoionisation modeller wishes to use, and the online application allows this in a convenient way.

The abundance scaling presented here is independent of the photoionisation model used. In any photoionisation modelling system, many parameters interact, and it is difficult to separate out the effects of a single parameter such as abundance scaling.  For this reason we leave detailed evaluation to a subsequent paper in this series that describes the latest revision of the $\textsc{Mappings}$ photoionisation code.

However, the effects of the new scaling can be significant, as illustrated in Figure \ref{fig1}.  Most important is the result of incorporating the primary and secondary nitrogen scaling. A problem with nebular models for a long time has been insufficient N$^+$ flux at higher metallicities.  Simple proportional scaling is incapable of producing sufficient flux in the key 6548 and 6584\AA\ [\rion{N}{2}] lines which are important in high redshift nebular observations.  The new non-linear scale appears to go a long way to remedying this problem, though more exploratory work is needed. Clearly, nitrogen abundance is complex, as it can be generated in  significant quantities early in Wolf-Rayet WN stars as well as later in AGB stars. Objects such as the Blue Compact Dwarf galaxy, HS0837+4717 \citep{2004AA...419..469P, 2011AA...532A.141P} have anomalously high nitrogen abundance at low oxygen abundance and must be modelled individually, rather than using a generic approach. Primary/secondary nitrogen scaling was modelled by \citet{2004ApJS..153....9G} in the context of AGNs, but is presented here more generally, as part of the new scaling relations with the revised standard abundance scale and fit parameters.

The new scaling presented here suggests that iron abundance is increasingly enhanced for 12+log(O/H) $>$ 8.26, compared to simple proportional scaling. This is important for modelling high excitation metal rich objects such as AGNs and Seyfert galaxies. Finally, the scaling techniques described here allow ready translation between the two major abundance scaling bases, iron and oxygen, used in stellar and nebular astrophysics.

\section*{Acknowledgments}
DCN and RSS thank Prof. Martin Asplund for helpful discussions and comments. DCN, MAD and LJK acknowledge the support of the Australian Research Council (ARC) through Discovery projects DP130103925 and DP160103631. BG acknowledges the support of the Australian Research Council (ARC) as the recipient of a Future Fellowship (FT140101202).

\appendix
\section{Minor nebular elements}

While Fe plays an important role in nebular cooling, the other iron-peak elements are of minor importance. Ni and Cl lines do appear in the spectra of \rion{H}{2} regions, but at very low flux. However, as we are arguing the need for common scaling in nebular, stellar atmosphere and stellar evolutionary track modelling, it is important to set out a basis for commonality in this work. In this appendix, for completeness, we present stellar data for Li, Be, B, F, Na, P, K, Sc, Ti, V, Cr, Mn, Co, Ni, Cu, and Zn.

\subsubsection{ Lithium, Beryllium and Boron}

These elements have very low abundances in stars and nebulae. Li is both produced and consumed in stars, so determining its primordial value from stellar spectra is difficult. It has been depleted in the sun by a factor of $\sim$150 compared to the meteoritic value \citep{2009ARAA..47..481A}, and Be and B have non-stellar primordial chemical evolution paths. Meteoritic values for Li, B and Be are acceptable as they are only trace elements. They play little role in nebular physics, so for the purposes of nebular modelling we assume they scale with oxygen.  Abundance values at the Galactic Concordance fiducial point are taken from the meteoritic [X/Si] values in \cite{2009LanB...4B...44L}, using 12+log(Si/H) = 7.533 derived from that paper as the conversion standard.

\subsubsection{Fluorine}

Solar photospheric abundance of fluorine can only be measured (indirectly) in sunspots from the HF abundances \citep[][and references therein]{2009ARAA..47..481A}. The solar values for fluorine from \cite{2009ARAA..47..481A} date back with minor variations to sunspot measurements, ca. 1970.  As fluorine is not normally observed in nebulae, in the Galactic Concordance we adopt the meteoritic values for fluorine from \cite{2009LanB...4B...44L}. This decision is not critical for nebular physics purposes, as fluorine does not play a significant role in the thermal balance of nebulae.

\subsubsection{Potassium, Scandium and Titanium}

\begin{figure*}
	\includegraphics[width=\textwidth]{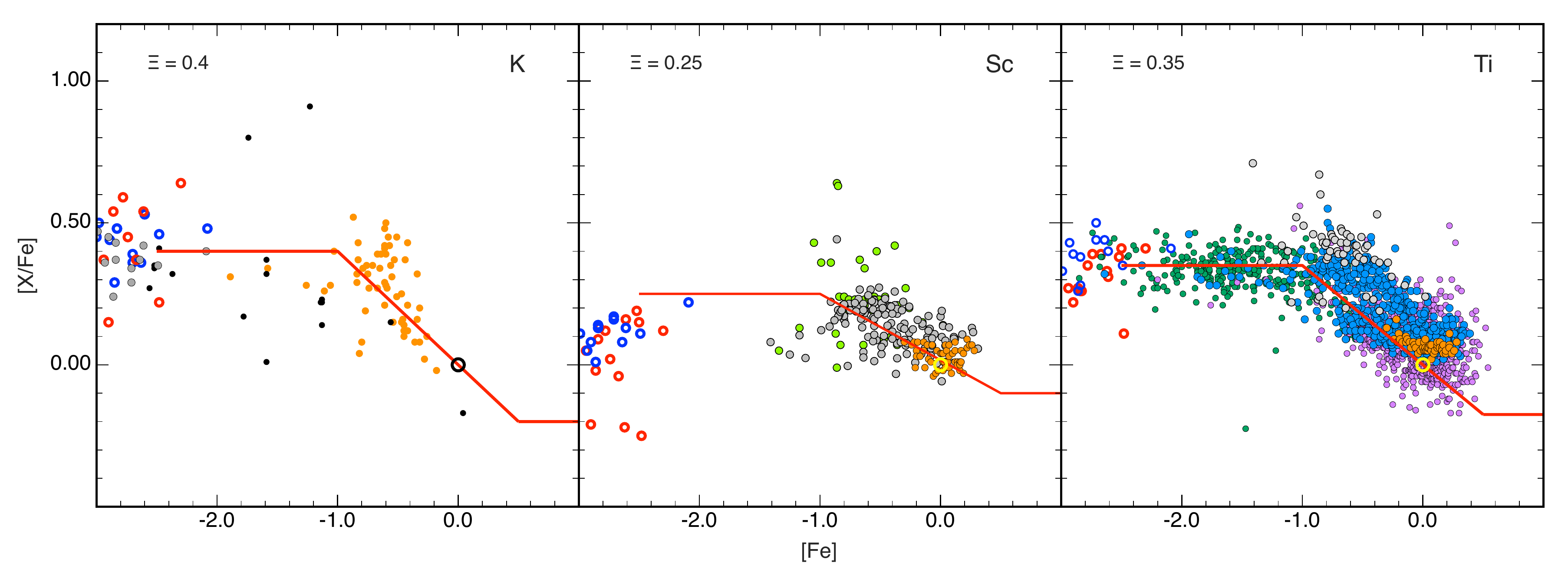}
    \caption{Scaling of K, Sc, Ti vs. Fe from stellar spectra. Sources:
    \citep[][K, orange]{2006AA...457..645Z},
    \citep[][K, black]{2009PASJ...61..563T},
    \citep[][K, grey]{2010AA...509A..88A},
    \citep[][K, Sc, Ti, blue circles]{2004AA...416.1117C},
    \citep[][K, Sc, Ti, red circles]{2015Natur.527..484H}
    \citep[][Sc, grey (thin disk), green (thick disk, halo)]{2012AA...545A..32A},
    \citep[][Sc, Ti, orange]{2013AA...552A...6G},
    \citep[][Ti, dark green]{2011ApJ...737....9R},
    \citep[][Ti, grey]{2012AA...545A..32A},
    \citep[][Ti, blue]{2014AA...562A..71B},
    \citep[][Ti, purple]{2014AJ....148...54H}
    }
    \label{figA1}
\end{figure*}

The elements K, Sc and Ti (Figure \ref{figA1}) also follow the $\alpha$-element trends.  Titanium and Scandium are sometimes included with the iron-peak elements, but its abundance shows a clear trend similar to the $\alpha$-elements. In other words, scandium and titanium appear to be generated in significant amounts in core-collapse supernovae.

\subsubsection{Sodium, Phosphorus and Zinc}

\begin{figure*}
	\includegraphics[width=\textwidth]{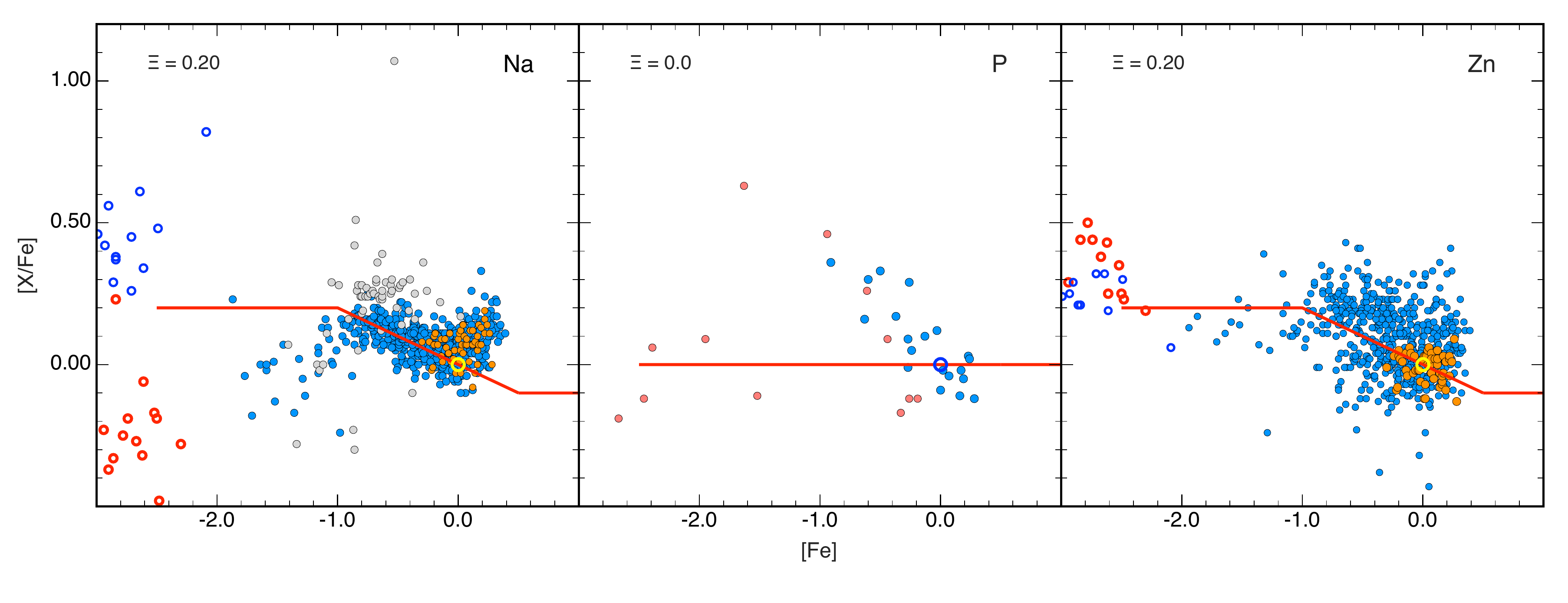}
    \caption{ Scaling of Na, P, Zn vs. Fe from stellar spectra. Sources:
    \citep[][Na, grey]{2012AA...545A..32A},
    \citep[][Na, Zn, orange]{2013AA...552A...6G},
    \citep[][Na, Zn, blue]{2014AA...562A..71B},
    \citep[][Na, Zn, red circles]{2015Natur.527..484H},
    \citep[][Na, Zn, blue circles]{2004AA...416.1117C},
    \citep[][P, blue]{2011AA...532A..98C},
    \citep[][P, orange]{2014ApJ...796L..24J}
    }
    \label{figA2}
\end{figure*}

Figure \ref{figA2} shows the behaviour of Na, P and Zn. The trends for these elements appear to be on the borderline between $\alpha$-group scaling and iron-peak scaling. The data for phosphorous \citep{2011AA...532A..98C, 2014ApJ...796L..24J} are so sparse that it is not possible to distinguish between the two apparent conflicting trends. Sodium appears to show a slight $\alpha$-process trend below [Fe/H] = 0. However, the upward trend at higher [Fe/H], previously attributed in open clusters to the ``Na/O anti-correlation''  observed in some globular clusters \citep{2012ApJ...756L..40G, 2012AA...548A.122B}, may be an artefact of the spectral data analysis as noted by  \citet{2015MNRAS.446.3556M}.

\subsubsection{Scaling iron-peak elements}

\begin{figure*}
    \includegraphics[width=1.0\textwidth]{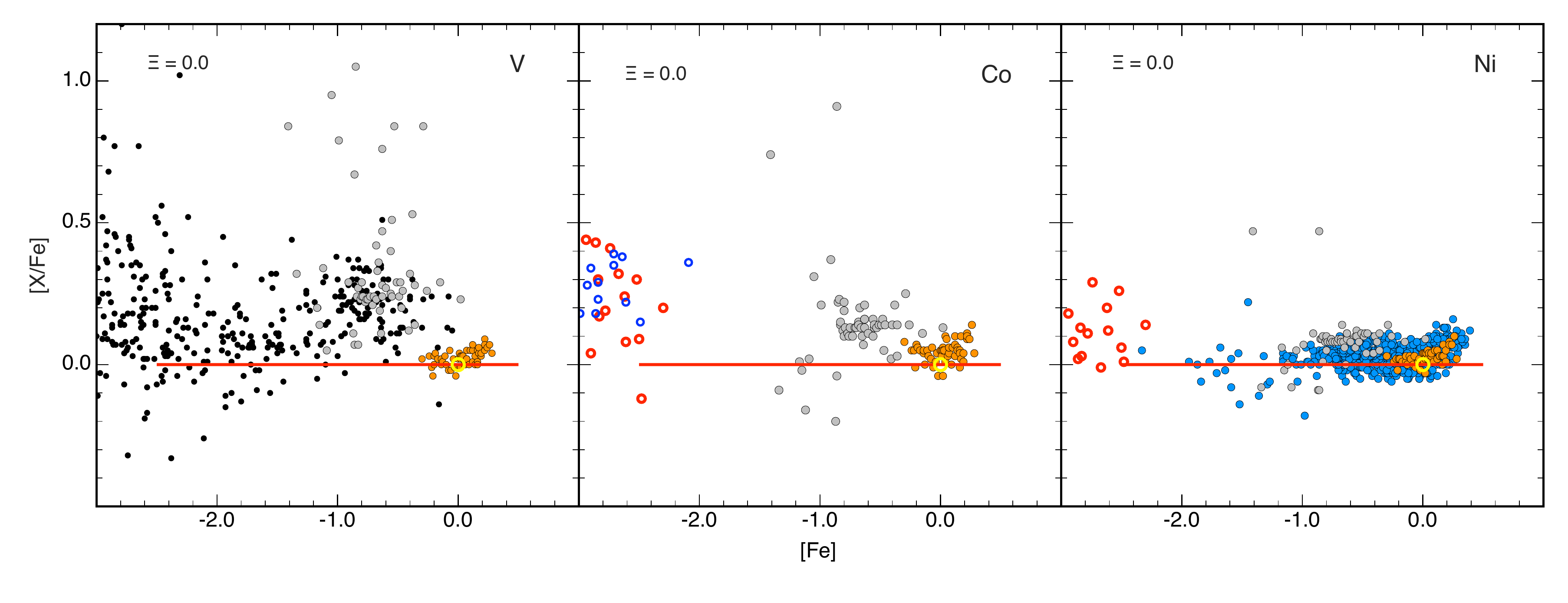}
    \caption{Scaling of V, Co, Ni vs. Fe from stellar spectra. Sources:
    \citep[][V, black]{2010AN....331..474F},
    \citep[][V, Co, Ni, grey]{2012AA...545A..32A},
    \citep[][V, Co, Ni, orange]{2013AA...552A...6G},
    \citep[][Ni, blue]{2014AA...562A..71B},
    \citep[][Co, Ni, red circles]{2015Natur.527..484H}
    \citep[][Co, blue circles]{2004AA...416.1117C}}
    \label{figA3}
\end{figure*}

\begin{figure*}
    \includegraphics[width=1.0\textwidth]{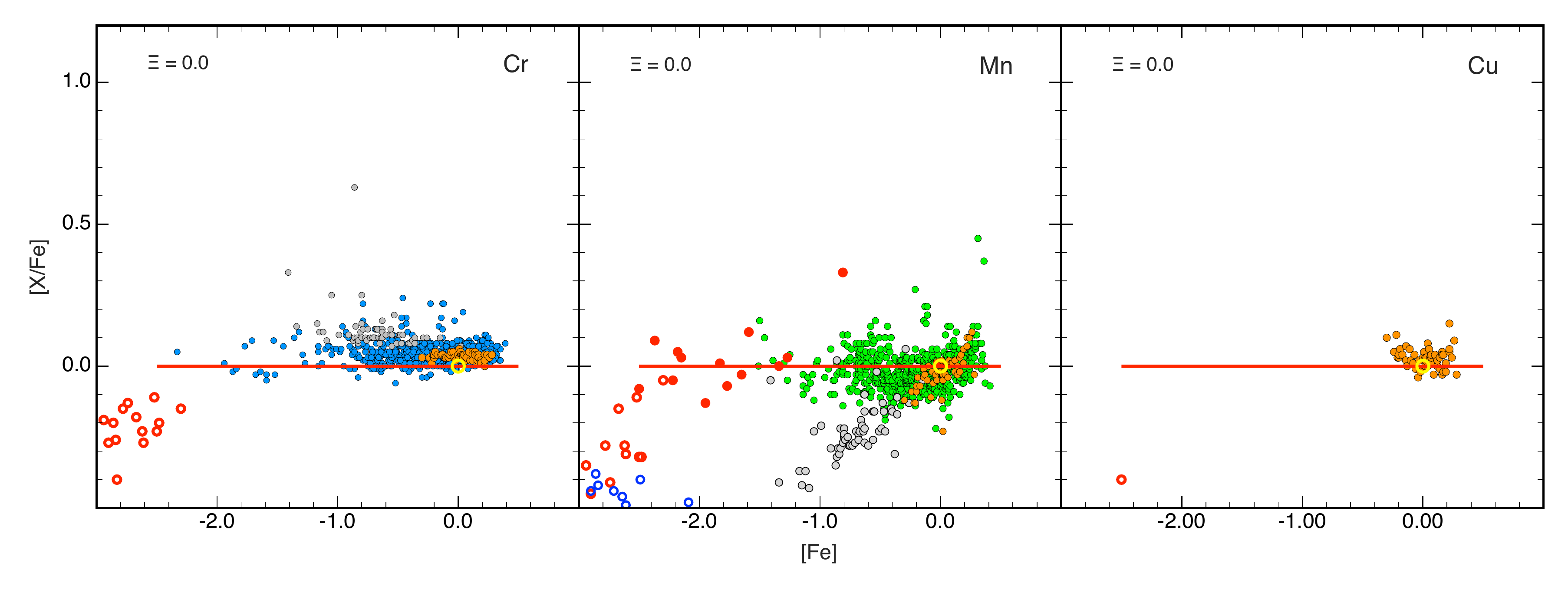}
    \caption{Scaling of Cr, Mn, Cu vs. Fe from stellar spectra. Sources:
    \citep[][Cr, Mn, grey]{2012AA...545A..32A},
    \citep[][Cr, Mn, Cu orange]{2013AA...552A...6G},
    \citep[][Cr, blue]{2014AA...562A..71B},
    \citep[][Mn, red disks]{2008AA...492..823B},
    \citep[][Mn, green]{2015AA...577A...9B},
    \citep[][Cr, Mn, Cu, red circles]{2015Natur.527..484H}
    \citep[][Mn, blue circles]{2004AA...416.1117C}. The trends to low values of [Cr/Fe] and [Mn/Fe] at low [Fe/H] may be real or artefacts, and are discussed in the text.}
    \label{figA4}
\end{figure*}

Iron-peak elements are produced in the nuclear fusion sequence that generates iron, so, at first glance, the abundances of vanadium, chromium, manganese, iron, cobalt, nickel and copper may be expected to remain constant as a fraction of iron. This is in large part supported by the stellar abundance data (Figures \ref{figA3} and \ref{figA4}), down to metallicities of [Fe/H] $\sim$-2.5.

However, at low metallicities, other trends are apparent in some elements. Decreasing trends below [Fe/H] $\sim$ -1.0 have been identified for chromium and manganese \citep[see, for example,][]{2004AA...416.1117C}. The abundances of chromium and manganese in Figure \ref{figA4} show an apparent decrease, and the abundance of cobalt (Figure \ref{figA3}) shows an apparent increase, at iron abundances [Fe/H] $\lesssim$ -2.5.

Some of the manganese data for [Fe/H] $>$ -2.0 may be an artefact of the spectral analysis, with NLTE effects not being taken into account.  In a sample of metal-poor stars, \citet{2008AA...492..823B} found that taking NLTE effects into account significantly increased the values obtained for [Mn/Fe], by up to 0.7dex, especially for lower iron abundances. Their data are shown in Figure \ref{figA4}, middle panel, as blue circles. \cite{2015AA...577A...9B} found the same effect. They calculated manganese abundances with and without NLTE effects, and the former follow the downward trend, whereas the data calculated taking into account NLTE physics remain constant with decreasing [Fe/H] (green circles, Figure \ref{figA4}, middle panel).  However, it is possible that the low Mn abundance trend at low [Fe] is real, as higher levels of Mn at [Fe] $<$ -1 are not well explained by core-collapse supernova models \citep{2013AA...559L...5S}. For nebular purposes this uncertainty is not significant, as Mn abundance has little effect in \rion{H}{2} region physics.

Similarly, for chromium, \citet{2010AA...522A...9B} found that neglecting NLTE effects in stellar modelling led to an apparent fall in chromium abundance at low metallicities. As NLTE effects are important in analysing the data \citep[as in the case of Na: see][]{2015MNRAS.446.3556M}, we reason that the abundances should scale directly with iron at least down to metallicities of [Fe/H] $\gtrsim$ -2.0.  Even if this is not the case, above [Fe/H] $\sim$ -2, the constant fit is acceptable to good.

Below [Fe/H] $\sim$ -2.5 the trends may be real, and were first reported for very low metallicity stars by \citet{1995AJ....109.2757M} and subsequently by \citet{1996ApJ...471..254R, 2001ApJ...561.1034N, 2004AA...416.1117C} and \citet{2013ApJ...762...26Y}. For example, \citet{1997ApJ...488..350N} found that high ($>$Z$_\odot$) carbon abundances ``are not uncommon'' for [Fe/H] $<$ -2.5 and suggest that this may may arise from different classes of core-collapse supernovae. It is possible that the disparate behaviour of other elements for [Fe/H] $< $ -2.5, may have the same origin. Because the purpose of the present work is to model metallicity behaviour for [Fe/H] $>$ -2, these effects do not affect the models presented here.

In the case of Cu, the data are lacking below [Fe/H] $\sim$ -0.4, but above that, they are constant, as shown in \cite[][Figure 2]{2013AA...552A...6G} which is an expanded plot of the data in Figure \ref{figA4}, right panel. When viewed at a larger scale for [Fe/H] $>$ -1.0, in some cases there appear to be slight upward trends in the iron-peak elements. This is apparent in the data from \citet[][their Figure 8]{2005AA...433..185B} \citet[][their Figure 16]{2014AA...562A..71B}, for example, but is not present in the data from \cite{2013AA...552A...6G}. It is not clear whether this is a real effect, or a result of abundance measurement methods.  The same trend appears in the sodium data (Figure \ref{figA2}, left panel), and, as noted above,  \citet{2015MNRAS.446.3556M} suggest this an artefact. Whatever the cause, it does not materially affect the fitting of a simple iron-peak constant trend with increasing [Fe/H]. While iron is critically important, especially at higher metallicities because of its abundance and presence in refractory dust, the other iron-peak elements are much less so.  For this reason, and the stellar abundance trends themselves, we are confident that assuming the iron-peak elements scale directly with iron is reasonable.

In general we have not included data from \citet{2010AN....331..474F} as they derive from several older sources for which the data reduction is uncertain. The exception is vanadium, where the lower metallicity data is scarce and the halo and thick disk data from \citet{2012AA...545A..32A} has a large spread, in order to clarify the likely behaviour at low metallicity.

\section{Additional tables}
This appendix contains a compendium of the major published standard solar abundances from 1976 to the present (Table \ref{tablea1}) a list of the stellar abundance sources we have used to derive the scaling fits (Table \ref{tablea2}) and the stellar populations used in the referenced studies (Table \ref{tablea3}).

\begin{landscape}
\begin{table*}
\scriptsize
\caption{Solar, meteoritic,  and B-star abundance standards}
\label{tablea1}
\begin{tabular}{rl*{15}c}
\hline
Z&Sym.& GC & SGA15 & NP12 & C11 & A09 & A09 & LPG09 & GAS07 & AGS05 & L03 & GS98 & AG89& AG89& AE82 & A76\\
\multicolumn{2}{c}{Element}&  &   &   &   &  proto. & photos. &  &  &  &  &  & photos. & metor. &  & \\
\hline
1 & H & 12.00 & 12.00 & 12.00 & 12.00 & 12.00 & 12.00 & 12.00 & 12.00 & 12.00 & 12.00 & 12.00 & 12.00 & 12.00 & 12.00 & 12.00\\
2 & He & 10.99 & \ldots & 10.99 & \ldots & 10.98 & 10.93 & 10.93 & 10.93 & 10.93 & 10.90 & 10.93 & 10.99 & 10.99 & 10.90 & 10.93\\
3 & Li & 3.28 & \ldots & \ldots & 1.03 & \ldots & 1.05 & 3.28 & 1.05 & 1.05 & 3.28 & 1.10 & 1.16 & 3.31 & 3.34 & 0.70\\
4 & Be & 1.32 & \ldots & \ldots & \ldots & \ldots & 1.38 & 1.32 & 1.38 & 1.38 & 1.41 & 1.40 & 1.15 & 1.42 & 1.46 & 1.10\\
5 & B & 2.81 & \ldots & \ldots & \ldots & \ldots & 2.70 & 2.81 & 2.70 & 2.70 & 2.78 & 2.55 & 2.60 & 2.88 & 2.95 & 3.00\\
6 & C & 8.42 & \ldots & 8.33 & 8.50 & 8.47 & 8.43 & 8.39 & 8.39 & 8.39 & 8.39 & 8.52 & 8.56 & 8.56 & 8.65 & 8.52\\
7 & N & 7.79 & \ldots & 7.79 & 7.86 & 7.87 & 7.83 & 7.86 & 7.78 & 7.78 & 7.83 & 7.92 & 8.05 & 8.05 & 7.96 & 7.96\\
8 & O & 8.76 & \ldots & 8.76 & 8.76 & 8.73 & 8.69 & 8.73 & 8.66 & 8.66 & 8.69 & 8.83 & 8.93 & 8.93 & 8.87 & 8.82\\
9 & F & 4.44 & 4.40 & \ldots & \ldots & \ldots & 4.56 & 4.44 & 4.56 & 4.56 & 4.46 & 4.56 & 4.56 & 4.48 & 4.49 & 4.60\\
10 & Ne & 8.09 & & 8.09 & \ldots & 7.97 & 7.93 & 8.05 & 7.84 & 7.80 & 7.87 & 8.08 & 8.09 & 8.09 & 8.11 & 7.92\\
11 & Na & 6.21 & 6.21 & \ldots & \ldots & \ldots & 6.24 & 6.29 & 6.17 & 6.17 & 6.30 & 6.33 & 6.33 & 6.31 & 6.32 & 6.25\\
12 & Mg & 7.56 & 7.59 & 7.56 & \ldots & 7.64 & 7.60 & 7.54 & 7.53 & 7.53 & 7.55 & 7.58 & 7.58 & 7.58 & 7.60 & 7.42\\
13 & Al & 6.43 & 6.43 & \ldots & \ldots & \ldots & 6.45 & 6.46 & 6.37 & 6.37 & 6.46 & 6.47 & 6.47 & 6.48 & 6.49 & 6.39\\
14 & Si & 7.50 & 7.51 & 7.50 & \ldots & 7.55 & 7.51 & 7.53 & 7.51 & 7.51 & 7.54 & 7.55 & 7.55 & 7.55 & 7.57 & 7.52\\
15 & P & 5.41 & 5.41 & \ldots & 5.46 & \ldots & 5.41 & 5.45 & 5.36 & 5.36 & 5.46 & 5.45 & 5.45 & 5.57 & 5.58 & 5.52\\
16 & S & 7.12 & 7.12 & \ldots & 7.16 & 7.16 & 7.12 & 7.16 & 7.14 & 7.14 & 7.19 & 7.33 & 7.21 & 7.27 & 7.28 & 7.20\\
17 & Cl & 5.25 & \ldots & \ldots & \ldots & \ldots & 5.50 & 5.25 & 5.50 & 5.50 & 5.26 & 5.50 & 5.50 & 5.27 & 5.29 & 5.60\\
18 & Ar & 6.40 & 6.40 & \ldots & \ldots & 6.44 & 6.40 & 6.50 & 6.18 & 6.18 & 6.55 & 6.40 & 6.56 & 6.56 & 6.58 & 6.80\\
19 & K & 5.04 & 5.04 & \ldots & 5.11 & \ldots & 5.03 & 5.11 & 5.08 & 5.08 & 5.11 & 5.12 & 5.12 & 5.13 & 5.14 & 4.95\\
20 & Ca & 6.32 & 6.32 & \ldots & \ldots & \ldots & 6.34 & 6.31 & 6.31 & 6.31 & 6.34 & 6.36 & 6.36 & 6.34 & 6.34 & 6.30\\
21 & Sc & 3.16 & 3.16 & \ldots & \ldots & \ldots & 3.15 & 3.07 & 3.17 & 3.05 & 3.07 & 3.17 & 3.10 & 3.09 & 3.09 & 3.22\\
22 & Ti & 4.93 & 4.93 & \ldots & \ldots & \ldots & 4.95 & 4.93 & 4.90 & 4.90 & 4.92 & 5.02 & 4.99 & 4.93 & 4.95 & 5.13\\
23 & V & 3.89 & 3.89 & \ldots & \ldots & \ldots & 3.93 & 3.99 & 4.00 & 4.00 & 4.00 & 4.00 & 4.00 & 4.02 & 4.04 & 4.40\\
24 & Cr & 5.62 & 5.62 & \ldots & \ldots & \ldots & 5.64 & 5.65 & 5.64 & 5.64 & 5.65 & 5.67 & 5.67 & 5.68 & 5.69 & 5.85\\
25 & Mn & 5.42 & 5.42 & \ldots & \ldots & \ldots & 5.43 & 5.50 & 5.39 & 5.39 & 5.50 & 5.39 & 5.39 & 5.53 & 5.54 & 5.40\\
26 & Fe & 7.52 & 7.47 & 7.52 & 7.52 & 7.54 & 7.50 & 7.46 & 7.45 & 7.45 & 7.47 & 7.50 & 7.67 & 7.51 & 7.52 & 7.60\\
27 & Co & 4.93 & 4.93 & \ldots & \ldots & \ldots & 4.99 & 4.90 & 4.92 & 4.92 & 4.91 & 4.92 & 4.92 & 4.91 & 4.92 & 5.10\\
28 & Ni & 6.20 & 6.20 & \ldots & \ldots & \ldots & 6.22 & 6.22 & 6.23 & 6.23 & 6.22 & 6.25 & 6.25 & 6.25 & 6.26 & 6.30\\
29 & Cu & 4.18 & 4.18 & \ldots & \ldots & \ldots & 4.19 & 4.27 & 4.21 & 4.21 & 4.26 & 4.21 & 4.21 & 4.27 & 4.28 & 4.50\\
30 & Zn & 4.56 & 4.56 & \ldots & \ldots & \ldots & 4.56 & 4.65 & 4.60 & 4.60 & 4.63 & 4.60 & 4.60 & 4.65 & 4.67 & 4.20\\\hline
\multicolumn{17}{l}{\textbf{Sources}}\\
\multicolumn{17}{l}{GC: this work}\\
\multicolumn{17}{l}{SGA15 solar photosphere: \cite{2015AA...573A..25S, 2015AA...573A..26S, 2015AA...573A..27G}}\\
\multicolumn{17}{l}{NP12 B stars: \cite{2012AA...539A.143N}}\\
\multicolumn{17}{l}{C11 solar photosphere: \cite{2011SoPh..268..255C}}\\
\multicolumn{17}{l}{A09: proto-solar and solar photosphere: \cite{2009ARAA..47..481A}}\\
\multicolumn{17}{l}{LPG09 meteoritic (and solar): \cite{2009LanB...4B...44L}}\\
\multicolumn{17}{l}{GAS07 solar photosphere: \cite{2007SSRv..130..105G}}\\
\multicolumn{17}{l}{AGS05 solar photosphere: \cite{2005ASPC..336...25A}}\\
\multicolumn{17}{l}{L03 meteoritic (and solar): \cite{2003ApJ...591.1220L}}\\
\multicolumn{17}{l}{GS98 solar photosphere: \cite{1998SSRv...85..161G}}\\
\multicolumn{17}{l}{AG89: solar  photosphere and meteoritic: \cite{1989GeCoA..53..197A}}\\
\multicolumn{17}{l}{AE82 solar photosphere: \cite{1982GeCoA..46.2363A}}\\
\multicolumn{17}{l}{A76 solar photosphere: \cite{1976asqu.book.....A}}\\
\hline
\end{tabular}
\end{table*}

\begin{table}
\footnotesize
\caption{Stellar data sources}
\label{tablea2}
\begin{tabular}{ll}
\\ \hline
Element & Sources\\
\hline
C  & \begin{minipage}[t]{\textwidth} \cite{1999AA...342..426G,2004AA...414..931A,2009AA...500.1143F,2005AA...430..655S,2012AA...539A.143N,2014AA...568A..25N} \end{minipage} \\
N  & \begin{minipage}[t]{\textwidth}\cite{2004AA...414..931A,2005AA...430..655S,2012AA...539A.143N} \end{minipage} \\
O  & \begin{minipage}[t]{\textwidth}\cite{2004AA...416.1117C,2011AA...535A..42T,2013AA...552A...6G,2013ApJ...764...78R,2014AA...562A..71B,2015MNRAS.454L..11A} \end{minipage} \\
Ne & \begin{minipage}[t]{\textwidth}\cite{2012AA...539A.143N} \end{minipage} \\
Na & \begin{minipage}[t]{\textwidth}\cite{2012AA...545A..32A,2013AA...552A...6G,2014AA...562A..71B,2015Natur.527..484H} \end{minipage} \\
Mg & \begin{minipage}[t]{\textwidth}\cite{2011ApJ...737....9R,2011AA...535A..42T,2012AA...545A..32A,2013AA...552A...6G,2014AA...562A..71B,2014AJ....148...54H,2015Natur.527..484H} \end{minipage} \\
Al & \begin{minipage}[t]{\textwidth}\cite{1993AA...275..101E,2004AA...416.1117C,2010AA...509A..88A,2013AA...552A...6G} \end{minipage} \\
Si & \begin{minipage}[t]{\textwidth}\cite{2011ApJ...737....9R,2011AA...535A..42T,2012AA...545A..32A,2013AA...552A...6G,2014AA...562A..71B,2014AJ....148...54H,2015Natur.527..484H} \end{minipage} \\
P  & \begin{minipage}[t]{\textwidth}\cite{2011AA...532A..98C,2014ApJ...796L..24J} \end{minipage} \\
S  & \begin{minipage}[t]{\textwidth}\cite{2011AA...528A...9S,2013AA...552A...6G} \end{minipage} \\
K  & \begin{minipage}[t]{\textwidth}\cite{2004AA...416.1117C,2006AA...457..645Z,2009PASJ...61..563T,2010AA...509A..88A,2015Natur.527..484H} \end{minipage} \\
Ca & \begin{minipage}[t]{\textwidth}\cite{2004AA...416.1117C,2011ApJ...737....9R,2011AA...535A..42T,2012AA...545A..32A,2013AA...552A...6G,2014AA...562A..71B,2015Natur.527..484H} \end{minipage} \\
Sc & \begin{minipage}[t]{\textwidth}\cite{2012AA...545A..32A,2013AA...552A...6G,2015Natur.527..484H} \end{minipage} \\
Ti & \begin{minipage}[t]{\textwidth}\cite{2004AA...416.1117C,2011ApJ...737....9R,2011AA...535A..42T,2012AA...545A..32A,2013AA...552A...6G,2014AA...562A..71B,2015Natur.527..484H} \end{minipage} \\
V  & \begin{minipage}[t]{\textwidth}\cite{2010AN....331..474F,2012AA...545A..32A,2013AA...552A...6G} \end{minipage} \\
Cr &\begin{minipage}[t]{\textwidth} \cite{2012AA...545A..32A,2013AA...552A...6G,2014AA...562A..71B,2015Natur.527..484H} \end{minipage} \\
Mn & \begin{minipage}[t]{\textwidth}\cite{2008AA...492..823B,2012AA...545A..32A,2013AA...552A...6G,2015AA...577A...9B,2015Natur.527..484H} \end{minipage} \\
Co & \begin{minipage}[t]{\textwidth}\cite{2012AA...545A..32A,2013AA...552A...6G,2015Natur.527..484H} \end{minipage} \\
Ni & \begin{minipage}[t]{\textwidth}\cite{2012AA...545A..32A,2013AA...552A...6G,2014AA...562A..71B,2015Natur.527..484H} \end{minipage} \\
Cu & \begin{minipage}[t]{\textwidth}\cite{2013AA...552A...6G} \end{minipage} \\
Zn & \begin{minipage}[t]{\textwidth}\cite{2013AA...552A...6G,2014AA...562A..71B,2015Natur.527..484H} \end{minipage} \\
\hline
 & See following table for stellar population details\\
\end{tabular}

\end{table}
\begin{table}
\centering
\caption{Stellar populations from cited sources}
\label{tablea3}
\begin{tabular}{ll}
\\ \hline
Source & Population(s)\\
\hline
\citet{1993AA...275..101E} & Nearby field disk dwarf F and G stars\\
\citet{1999AA...342..426G} & as above \\
\citet{2004AA...414..931A} & Metal-poor halo stars \\
\citet{2004AA...416.1117C} & Very metal-poor dwarfs and giants \\
\citet{2005AA...430..655S} & Halo unmixed extremely metal-poor giants \\
\citet{2006AA...457..645Z} & Nearby metal-poor stars \\
\citet{2008AA...492..823B} & Halo stars \\
\citet{2009AA...500.1143F} & Halo stars \\
\citet{2009PASJ...61..563T} & Red giants in metal-poor globular clusters \\
\citet{2006AA...457..645Z} & Nearby metal-poor stars \\
\citet{2008AA...492..823B} & Halo stars \\
\citet{2009AA...500.1143F} & Halo stars \\
\citet{2010AA...509A..88A} & Disk and halo stars \\
\citet{2010AN....331..474F} & Metal-poor and extremely metal-poor stars from several older sources \\
\citet{2011AA...528A...9S} & Extremely metal-poor stars \\
\citet{2011AA...532A..98C} & Cool galactic disk stars \\
\citet{2011ApJ...737....9R} & Thick disk metal-poor stars \\
\citet{2012AA...539A.143N} & Nearby thick disk B-stars \\
\citet{2012AA...545A..32A} & Thin disk, thick disk and halo \\
\citet{2013AA...552A...6G} & F- and G-type main sequence stars \\
\citet{2014AA...562A..71B} & F and G dwarf and sub-giant stars in the solar neighbourhood \\
\citet{2014AA...568A..25N} & F and G main sequence stars in the solar neighbourhood \\
\citet{2014AJ....148...54H} & Stars within 150pc of the Sun \\
\citet{2014ApJ...796L..24J} & Low metallicity stars \\
\citet{2015AA...577A...9B} & Thick and thin disk stars \\
\citet{2015MNRAS.454L..11A} & Dwarfs and subgiants \\
\citet{2015Natur.527..484H} & Extremely metal-poor stars in the Milky Way bulge \\
\hline
\end{tabular}
\end{table}

\end{landscape}

\end{document}